\documentclass[a4paper,12pt]{article}

\usepackage{amssymb,amsfonts,amstext,mathrsfs}
\usepackage[sumlimits,intlimits,namelimits]{amsmath}
\usepackage{icomma}
\usepackage{bbold}
\usepackage[english]{babel}
\usepackage[utf8]{inputenc}
\usepackage{graphicx}
\usepackage{color}
\usepackage{tabularx}
\usepackage{booktabs}
\usepackage{array}
\usepackage{float}
\restylefloat{figure}
\usepackage{ulem}
\usepackage{hyperref}
\usepackage{units}
\usepackage{cancel}
\usepackage{tikz}
\usepackage{xcolor}
\usepackage{slashed}
\usetikzlibrary{arrows,shapes,trees}

\usepackage{cite}

\usepackage{cancel} 
\usepackage{extarrows}
\usepackage{geometry}
\usepackage{subfigure}
\newcommand{\re}[1]{\mathbf{#1}}
\newcommand{\rb}[1]{\mathbf{\overline{#1}}}
\newcommand{\VEV}[1]{\langle#1\rangle}
\newcommand{\eV}{\mathrm{eV}}
\newcommand{\MeV}{\mathrm{MeV}}
\newcommand{\GeV}{\mathrm{GeV}}
\newcommand{\TeV}{\mathrm{TeV}}

\newcommand{\id}{\mathbb{1}}
\newcommand{\hc}{{h.c.}}
\newcommand{\be}{\begin{equation}}
\newcommand{\ee}{\end{equation}}
\newcommand{\bea}{\begin{eqnarray}}
\newcommand{\eea}{\end{eqnarray}}

\newcommand{\cleqn}{\setcounter{equation}{0}}

\begin{document}

\title{\hfill ~\\[-30mm]
          \hfill\mbox{\small  QFET-2015-01}\\[-3.5mm]
          \hfill\mbox{\small  SI-HEP-2015-01}\\[13mm]
       \textbf{
Combining Pati-Salam and \\[2mm] Flavour Symmetries
}}

\author{
Thorsten Feldmann\footnote{E-mail: {\tt thorsten.feldmann@uni-siegen.de}}
,\,
Florian Hartmann\footnote{E-mail: {\tt hartmann@physik.uni-siegen.de}}
,\,\\[2mm]
Wolfgang Kilian\footnote{E-mail: {\tt kilian@physik.uni-siegen.de}}
,\,
Christoph Luhn\footnote{E-mail: {\tt christoph.luhn@uni-siegen.de}}\\[3mm]
\it{\small Theoretische Physik 1, Naturwissenschaftlich-Technische
    Fakult\"at,}\\ \it{\small Universit\"at Siegen,
 57068 Siegen, Germany}}

\date{\today}

\maketitle

\begin{abstract}
\noindent  
We construct an extension of the Standard Model (SM)
which is based on grand unification with Pati-Salam symmetry.
The setup is supplemented with the idea of spontaneous flavour symmetry
breaking which is mediated through flavon fields with renormalizable 
couplings to new heavy fermions.
While we argue that the new gauge bosons in this approach can be
sufficiently heavy to be irrelevant at low energies, the fermionic
partners of the SM quarks, in particular those for the third generation,
can be relatively light and provide new sources of flavour violation.
The size of the effects is constrained by the observed values of the
SM Yukawa matrices, but in a way that is different from the standard
minimal-flavour violation approach. 
We determine characteristic deviations from the SM that could eventually 
be observed in future precision measurements.
\end{abstract}
\thispagestyle{empty}
\vfill

\newpage 
\setcounter{page}{1}




\section{Introduction}
\cleqn

Grand Unified Theories (GUTs) represent an attractive direction to search for 
extensions of the Standard Model (SM) of particle physics (see
e.g.~\cite{Buras:1977yy}).
The peculiar structure of the SM gauge group and its 
fermionic representations observed at low energies finds a more
natural-looking explanation when embedded into larger (but simpler)
groups that may be realized at higher energies. In particular, 
each generation of SM quarks and leptons fits into a 16-dimensional spinor 
representation of the group $SO(10)$~\cite{Fritzsch:1974nn} 
which then also contains a right-chiral neutrino that can be used to explain
the tiny but non-vanishing neutrino masses via the seesaw
mechanism~\cite{Minkowski:1977sc,GellMann:1980vs,Yanagida:1979as,Mohapatra:1979ia}.
Theoretical complication re-enters when it comes to the question of how the
GUT symmetry is broken down to the $SU(3)\times SU(2)\times U(1)$ symmetry of
the SM. Different Higgs representations of the GUT group can be chosen, and 
the symmetry breaking further depends on the parameters controlling the ground
state of the scalar potential. It turns out that the SM group can be reached
from $SO(10)$ via basically two separate routes: either via the Georgi-Glashow
group $SU(5)$~\cite{Georgi:1974sy}, or via the Pati-Salam (PS) group  
$SU(4) \times SU(2) \times SU(2)'$~\cite{Pati:1973uk,Pati:1974yy}.

Adopting a non-supersymmetric framework, we will focus on the PS gauge group
with an explicit $Z_2$ symmetry relating the couplings and representations of
the two $SU(2)$ group factors (which affect the left- and right-chiral SM
fermions, respectively). 
Recent work in this framework~\cite{Hartmann:2014fya,Kilian:2006hh,Howl:2007hq,
Calibbi:2009cp,Braam:2009fi,DeRomeri:2011ie,Arbelaez:2013hr,Arbelaez:2013nga}
suggests that it is worthwhile to consider PS scenarios with an extended field
content (as compared to minimal setups) where the breaking from the PS group
to the SM group involves several distinct mass scales which eventually realize
gauge-coupling unification at the GUT scale. Here we investigate a particular
construction which also allows us to address questions relating to flavour and
flavour mixing. 

The idea of combining grand unification with flavour is itself not new (for reviews see
e.g.~\cite{Chen:2003zv,King:2003jb,Mohapatra:2006gs,Altarelli:2010gt,King:2013eh,King:2014nza}). 
The intimate connection of the SM fermions, which is dictated by the GUT
multiplet structure, generally relates the Yukawa matrices of different
sectors in an unrealistic way. In order to accommodate the observed Yukawa
structures, it is therefore necessary to introduce more than one GUT-breaking
Higgs field~\cite{Georgi:1979df,Antusch:2009gu}. Flavour symmetries are then 
imposed to control the resulting Yukawa patterns by suitably chosen vacuum
expectation values (VEVs) of scalar flavon fields. Typically, the
so-constructed Yukawa matrices arise as sums of mostly non-renormalizable
terms~\cite{Froggatt:1978nt,Leurer:1992wg,Leurer:1993gy}. For explicit
constructions compatible with the PS gauge group, see e.g.~\cite{
King:2003rf,
King:2006me
,King:2006np
,King:2009mk
,Dutta:2009bj
,King:2009tj
,Toorop:2010yh
,deMedeirosVarzielas:2011wx
,King:2014iia
,Hartmann:2014ppa}.

In contrast to this traditional approach, 
we implement an idea first proposed 
by Grinstein, Redi and Villadoro (GRV) in~\cite{Grinstein:2010ve}, where 
the concept of gauged flavour symmetries is realized within a renormalizable
field theory. Among others, this requires to introduce new heavy partners 
for the SM fermions in order to cancel gauge anomalies associated 
with the chiral representations of the flavour symmetry 
group (see also~\cite{Albrecht:2010xh}).
The new fermions can then be used to mediate flavour-symmetry breaking 
---
which is realized by VEVs of new matrix-valued
scalar flavon fields\footnote{The possibility of obtaining hierarchical
structures for flavon VEVs (and thus realizing sequential flavour-symmetry
breaking as described in~\cite{Feldmann:2009dc}) by an appropriate
flavour-invariant potential has been discussed, for instance,
in~\cite{Alonso:2011yg,Nardi:2011st,Alonso:2012fy,Espinosa:2012uu,Alonso:2013mca,Alonso:2013nca,Alonso:2013dba,Fong:2013dnk,Merlo:2015nra}.} 
--- 
to the SM sector. The original GRV paper focused on the SM quark multiplets.
With a minimal set of new fermions and scalars, the authors have shown that the 
SM Yukawa matrices $Y_{u,d}$ for up- and down-type quarks are effectively 
given by the inverse of the corresponding flavon VEVs. 
Here, the partner of the top quark $t'$ plays a special role: 
the large Yukawa coupling of the top quark implies that $t'$ can be relatively
light (i.e.\ within the reach of direct searches at the 
LHC at CERN), and that it can have a substantial mixing with the SM top quark.
This leads to interesting new physics effects in electroweak precision
observables and flavour transitions involving the third generation of quarks. 
At the same time, indirect flavour bounds from the first and second generation
(most notably from $K-\overline{K}$ mixing and rare kaon decays) are naturally
satisfied (see also~\cite{Buras:2011wi}). 
Variants of the GRV idea in the context of the SM~\cite{D'Agnolo:2012ie},
GUTs~\cite{Feldmann:2010yp,Guadagnoli:2011id,Mohapatra:2012km}, 
and supersymmetric theories~\cite{Krnjaic:2012aj,Franceschini:2013ne} have
also been discussed in the literature. 

The new essential ingredients in this work are a consequence of embedding the
GRV mechanism in an explicit left-right (i.e.\ $Z_2$) symmetric PS GUT with a Higgs
bi-doublet. First, this requires to introduce (at least) two partners for each
conventional PS fermion multiplet. 
Second, in order to obtain realistic masses and mixing for quarks and leptons,
one has to introduce different types of flavon fields. In the minimal setup that
we will construct for the quark sector, we consider flavons that transform as
singlets and triplets under the $SU(2)$ factors of the PS symmetry group. The
breaking of the flavour symmetry by the singlet VEV alone treats up- and
down-type quarks in the same way, such that the corresponding Yukawa matrices
are identical, $Y_u = Y_d$. An additional VEV for the $SU(2)'$ triplet flavon
then guarantees that $Y_u \neq Y_d$. We also discuss how this framework can be
extended to yield realistic masses and mixings for the charged leptons
and neutrinos.

Our paper is organized as follows. In Section~\ref{sec:generalsetup}, we
introduce the general setup of our PS flavour theory and discuss the various
scales of symmetry breaking. Moreover, we deduce approximate expressions for
the Yukawa matrices as well as the effective light neutrino mass matrix.
Section~\ref{sec:GRVQuarkFlavour} focuses on the flavour
structure of the quark sector. We derive analytic formulas for the
transformations from the flavour to the mass basis in order to extract the 
(non-standard) couplings of the quarks to the SM gauge bosons and the
Higgs. In Section~\ref{sec:scan}, we describe the method we have used to scan
over the parameter space of the model. The results of this numerical scan are
presented and discussed in detail in Section~\ref{sec:GRVQuarkPheno}. We
conclude in Section~\ref{sec:conclus}.
Supplementary material and more technical details are provided in the Appendix.




\section{\label{sec:generalsetup}Pati-Salam model with gauged flavour symmetry}
\cleqn

As outlined in the introduction, our starting point is a non-supersymmetric
high-energy theory with an underlying Pati-Salam gauge symmetry.
A manifest left-right symmetric setup is realized by enforcing a discrete $Z_2$ symmetry. 
The flavour symmetry group in PS is defined by the independent transformations 
of the two fermion representations which are used to embed the SM fermions.
In the following, we restrict ourselves to the flavour group $SU(3)_I\times
SU(3)_{II}$, where the left-chiral SM fermions transform as triplets under
$SU(3)_I$ while the right-chiral SM fermions furnish triplet representations
of $SU(3)_{II}$. This flavour symmetry, which we assume to be gauged, is
spontaneously broken by the VEVs of flavon fields. By the usual Higgs mechanism, 
the associated Goldstone modes will become the longitudinal degrees of freedom of 
sixteen new heavy gauge bosons of the flavour group.\footnote{The Yukawa
sector is additionally invariant under three independent $U(1)$
symmetries: two associated with $SU(3)_I\times SU(3)_{II}$, while the third
relates to the neutrino extension of the theory. All three are broken
spontaneously by flavons of the  neutrino sector. Massless
Goldstone modes can, however, be avoided by breaking the $U(1)$ symmetries
explicitly in the scalar potential, for instance by terms involving the determinant
of flavon fields. 
}
Notice that the maximal flavour-symmetry group in PS is smaller than in the SM, 
where we encounter a $U(3)^3$ symmetry in the quark sector 
(as discussed in the original GRV paper \cite{Grinstein:2010ve}) and 
a $U(3)^2$ symmetry in the lepton sector.
To summarize, the full symmetry of the Lagrangian is given by  
\begin{align}\label{eq:symDEF}
\mathcal G~=~\underbrace{\Big( SU(4)\times SU(2)\times SU(2)' \Big)}_{\text{Pati-Salam}}\;\times\; 
\underbrace{\Big( SU(3)_{I}\times SU(3)_{II} \Big)}_{\text{flavour}}
    \;\times\; Z_2\ ,
\end{align}
where the global $Z_2$ symmetry is realized such that a general representation 
$\Omega$ of $\mathcal G$ is transformed into $\widetilde{\Omega}$ of 
$\mathcal G$ as
\begin{align}
\label{eq:z2}
 \Omega= (\omega_c,\,\omega,\,\omega')(\omega_{I},\,\omega_{II}) 
 \,\quad\xlongrightarrow{~~Z_2~~} \quad\,\widetilde{\Omega} = 
(\overline{\omega}_c,\,\omega',\,\omega)(\overline{\omega}_{II},\,\overline{\omega} 
_{I})\ .
\end{align}
Here and in the following, $\omega$ denotes a representation of the
corresponding group factor and $\overline \omega$  is its complex conjugate. As we
will only introduce (pseudo-)real representations of $SU(2)\times SU(2)'$, we
have dropped the complex conjugation of the corresponding representations in
$\widetilde \Omega$ on the right-hand side of Eq.~\eqref{eq:z2}.


\subsection{Particle content and renormalizable Lagrangian} 

The left- and right-chiral SM fermions $q_L$ and $q_R$ (including the right-chiral
neutrino) are embedded in the $(\re4,\re2,\re1)(\re3,\re1)$ 
and $(\re4,\re1,\re2)(\re1,\re3)$ representations of $\mathcal G$. Notice that
the transformation of Eq.~\eqref{eq:z2} maps the right-chiral fermions $q_R$
into the complex conjugate $\overline q_L$ of the left-chiral fermions. 
Therefore, the chirality of the spinors remains unchanged under a $Z_2$
transformation. The SM Higgs is embedded in the bi-doublet $H \sim
(\re1,\re2,\re2)(\re1,\re1)$, which effectively generates the structure of a
two-Higgs-doublet model. 
In order to generalize the idea of GRV, it is necessary to introduce a set of
new  fermions, denoted as $\Sigma_{L,R}$, $\Xi_{L,R}$. The former,
i.e.\ $\Sigma_L$ and $\Sigma_R$, transform as $(\re4,\re1,\re2)(\re1,\re3)$ and
$(\re4,\re1,\re2)(\re3,\re1)$, while the complex conjugate of the latter,
i.e.\ $\overline{\Xi}_{L,R}$, correspond to the $Z_2$ conjugated
representations of $\Sigma_{R,L}$. Since we impose an  exact $Z_2$ symmetry,
we have to introduce both pairs.

Due to the single Higgs bi-doublet~$H$, a  flavour symmetry breaking
flavon field~$S$ transforming trivially under PS cannot discriminate between
the up-type and the down-type sector. In contrast to the original GRV
model~\cite{Grinstein:2010ve}, it is 
therefore impossible to generate a non-trivial flavour structure without
mediating the $SU(2)'$ breaking of the PS sector to the flavour sector.
As a consequence, we are led to introduce PS non-singlet flavon fields,
where the simplest case is realized by
$T\sim (\re1,\re3,\re1)(\rb3,\re3)$ and $T' \sim (\re1,\re1,\re3)(\rb3,\re3)$, 
in addition to the singlet flavon~$S$.

\begin{table}[t]
\centering
 \begin{tabular}{lccc}
   \toprule 
   ~~~~~~~ 
& Pati-Salam Symmetry & Flavour Symmetry & VEV\\  
& ~$SU(4)\times SU(2)\times SU(2)'$~ & $SU(3)_{I}\times SU(3)_{II}$ &  \\
   \midrule
   ~$\overline{q}_L$ & $(\rb4\,,\,\re2\,,\,\re1)$ & $(\rb3\,,\,\re1)$ &--- \\
   ~$q_R$ & $(\re4\,,\,\re1\,,\,\re2)$& $(\re1\,,\,\re3)$ &--- \\
   ~$H$  & $(\re1\,,\,\re2\,,\,\re2)$ & $(\re1\,,\,\re1)$ & $v_u$,$v_d$\\[1ex]
   ~$\overline{\Sigma}_L$ & $(\rb4\,,\,\re1\,,\,\re2)$ & $(\re1\,,\,\rb3)$ &--- \\
   ~$\Sigma_R$ & $(\re4\,,\,\re1\,,\,\re2)$ & $(\re3\,,\,\re1)$ &---  \\
   ~$\overline{\Xi}_L$ & $(\rb4\,,\,\re2\,,\,\re1)$ & $(\re1\,,\,\rb3)$ &--- \\
   ~$\Xi_R$ & $(\re4\,,\,\re2\,,\,\re1)$ & $(\re3\,,\,\re1)$ & ---\\[1ex]
   ~$T$ & $(\re1\,,\,\re3\,,\,\re1)$ & $(\rb3\,,\,\re3)$ & 0 \\
   ~$T'$ & $(\re1\,,\,\re1\,,\,\re3)$ & $(\rb3\,,\,\re3)$ & $\pm t' \,M$ \\
   ~$S$ & $(\re1\,,\,\re1\,,\,\re1)$ & $(\rb3\,,\,\re3)$ & $s\,M$ \\\midrule
   ~$\overline\Theta_L$ & $(\re1\,,\,\re1\,,\,\re1)$ & $(\rb3\,,\,\re8)$ & ---\\
   ~$\Theta_R$ & $(\re1\,,\,\re1\,,\,\re1)$ & $(\re8\,,\,\re3)$ & ---\\[1ex]
   ~$S_\nu$ & $(\re1\,,\,\re1\,,\,\re1)$ & $(\re6\,,\,\re1)$ & $s_\nu \Lambda_\nu$\\
   ~$S'_\nu$ & $(\re1\,,\,\re1\,,\,\re1)$ & $(\re1\,,\,\rb6)$  & $s'_\nu \Lambda_\nu$ \\[1ex]
   ~$\Phi$ & $(\re4\,,\,\re2\,,\,\re1)$ & $(\re8\,,\,\re1)$ & 0\\
   ~$\Phi'$ & $(\rb4\,,\,\re1\,,\,\re2)$ & $(\re1\,,\,\re8)$ & $\varphi'\Lambda_\varphi$ \\
   \bottomrule
 \end{tabular}
\caption{The particle content of the theory with imposed Pati-Salam and
  flavour symmetry. Left- and right-chiral fermions $\psi_{L,R}$ are denoted
  by subscripts $L$ and $R$, respectively. The VEVs of the scalar fields (no subscript $L$ or
  $R$) are given in the rightmost column. The lower part of the table shows fields
  necessary for generating Majorana neutrino masses.} 
\label{tab:fullfieldcontent}
\end{table}

The particle content defined up to this point treats quarks and leptons on
equal footing. In particular, neutrinos would be Dirac particles which copy
the hierarchical pattern of the up-type quarks. In order to account for
the evident difference of quark and lepton flavour, we take advantage
of the possibility to generate Majorana mass terms for electrically
neutral leptons.
Demanding renormalizability, it is necessary to extend the particle content by
further fermionic and scalar fields. Although not unique, there exists a
preferred extension (see Appendix~\ref{app:leptons}) in which additional
PS-neutral fermions $\overline \Theta_{L}$ and $\Theta_{R}$ acquire Majorana
masses via their coupling to new flavour symmetry breaking flavons $S_\nu$ and
$S'_\nu$. Furthermore, $\overline \Theta_L$ and $\Sigma_R$ ($\Theta_R$ and 
$\overline \Xi_L$) are coupled to one another by means of yet another flavon
field~$\Phi'$ ($\Phi$). Transforming as $(\rb4,\re1,\re2)$ under PS, the 
vacuum expectation value of the flavon $\Phi'$ breaks the Pati-Salam
gauge symmetry in the direction of the SM singlet, thereby projecting out the
SM neutral component of $\Sigma_R$. This in turn couples to the neutrino
component of $\overline q_L$ (either directly or indirectly via
$\overline{\Sigma}_L$ and~$q_R$) which
eventually generates a Majorana mass for the left-chiral neutrino $\overline
q_L^\nu$.  

The particle content of the model required for the description of
quark and lepton masses and mixing is summarized in
Table~\ref{tab:fullfieldcontent}, together with the corresponding
transformation properties under the underlying PS and flavour symmetry.
Note that we list all fermions using only right-chiral degrees of freedom,
i.e.\ $\psi_R$ and the Dirac adjoint~$\overline \psi_L$ (instead of $\psi_L$).
As each particle is accompanied by its $Z_2$ conjugate partner, the PS anomaly
$[SU(4)]^3$ vanishes by construction. Concerning the flavour anomalies
$[SU(3)_I]^3$ and $[SU(3)_{II}]^3$, it is straightforward to check that 
the number of fermionic triplets matches the number of fermionic
anti-triplets. Notice that we have chosen the octet representation in
$\overline \Theta_L$ and $\Theta_R$ in order to reflect the dimension of the
PS representation of the other fermions. As the adjoint does not contribute to
the $SU(3)$ anomaly, both flavour anomalies vanish identically. 
Mixed anomalies do not exist due to the absence of $U(1)$ factors in $\mathcal
G$ of Eq.~\eqref{eq:symDEF}. Hence the model presented in
Table~\ref{tab:fullfieldcontent} is free of any gauge anomaly.


With the particle content specified in Table~\ref{tab:fullfieldcontent}, it is
straightforward to formulate the most general renormalizable Yukawa Lagrangian
which is invariant under the full symmetry~$\mathcal G$. Collecting all terms
which involve the heavy fermions $\Theta_{L,R}$ (required to generate light
Majorana neutrino masses) in $\mathcal L^\nu_{\text{Yuk}}$, we can write 
\be
\mathcal L_{\text{Yuk}}~=~ \mathcal L^q_{\text{Yuk}} \,+\,\mathcal
L^\nu_{\text{Yuk}} \ ,\label{eq:lagrangian}
\ee
where 
  \begin{align}
\label{eq:fermionlagrangian}
     \mathcal{L}^q_\text{Yuk} \,= 
     &\phantom{+}\; \lambda\;\overline{q}_{L}\,H\,\Sigma_{R}
     + \overline{\Sigma}_L\,\left(\kappa_S\,S+\kappa_T\,T'\right)\,\Sigma_R
     + M \,\overline{\Sigma}_L\, q_{R} \;+h.c. \cr 
     &+\lambda\;\overline{\Xi}_{L}\,H\,q_R
     + \overline{\Xi}_L\,\left(\kappa_S\,S+\kappa_T\,T\right)\,\Xi_R
     + M \,\overline{q}_L\,\Xi_{R} \;+h.c. \ ,
\end{align}
and
\begin{align}
\label{eq:fermionlagrangian-nu}
     \mathcal{L}^\nu_\text{Yuk} \,\sim 
     &\phantom{+}\:\overline{\Theta}_{L}\,\Phi'\,\Sigma_{R}
     +\mbox{$\frac{1}{2}$} \, \overline{\Theta}_L\, \,S_\nu\,\overline\Theta_L \;+h.c. \cr 
      &+\overline{\Xi}_{L}\,\Phi\,\Theta_{R}\hspace{0.25mm}
     + \mbox{$\frac{1}{2}$} \,{\Theta}_R\, \,S'_\nu\,\Theta_R \,+h.c. \hspace{42.85mm} \cr 
     &+\overline{\Theta}_{L}\,S^\dagger \,\Theta_{R}  \;+h.c. \ .
  \end{align}
In Eqs.~\eqref{eq:fermionlagrangian} and~\eqref{eq:fermionlagrangian-nu},  
the terms of the first line are related to those of the second line through
the action of the $Z_2$ symmetry. The term in the third line of 
Eq.~\eqref{eq:fermionlagrangian-nu} is its own $Z_2$ conjugate.
Due to these relations, the Yukawa Lagrangian $\mathcal L^q_{\text{Yuk}}$
depends on only four independent parameters: three dimensionless couplings
$\lambda$, $\kappa_S$, $\kappa_T$, and one mass scale~$M$.
Redefining the phases of the fermions, it is clear that $\lambda$ and $M$
can be chosen real and positive without loss of generality. As $\kappa_S$ and
$\kappa_T$ appear only in combination with the flavon fields which will generally
acquire complex VEVs, their phases can be kept arbitrary at this point.
With regard to the neutrino Lagrangian $\mathcal{L}_\text{Yuk}^\nu$,
we are at this point only interested in a qualitative analysis. Therefore,
we have suppressed explicit coupling constants, which would again be related
by the $Z_2$ symmetry, in Eq.~\eqref{eq:fermionlagrangian-nu}.


\subsection{Symmetry breaking and mass scales}
\label{sec:SBandMS}

The flavour structure of the SM fermions originates in the flavon
fields. Assigning VEVs of order $M$ or higher for the components of $S$ and $T'$, breaks the
flavour symmetry $SU(3)_I \times SU(3)_{II}$ above the electroweak
scale~$v$. The scalar field $T$, on the other hand, must not get a high scale VEV
as it transforms as an $SU(2)$ triplet. However, a possible mixing with the
electroweak Higgs doublets would generically induce a VEV $\VEV{T}$ below the
electroweak scale at order~$v^2/M$. As the impact of this VEV on the flavour
structure is negligible compared to~$\VEV S$, we will approximate $\VEV T\simeq 0$
in the following, for simplicity. In order to obtain a phenomenologically realistic difference
between the effective Yukawa matrices of the up and the down quarks, we need the singular values of 
the VEVs $\VEV{S}$ and $\VEV{T'}$ to be of similar size. 

In addition to breaking the flavour symmetry, the VEV $\VEV{T'}$ breaks
$SU(2)'$ as well. In order to keep the standard charge assignments for up- and
down-type quarks, we choose $\VEV{T'}$  to be aligned along the $\tau_3$ direction
of $SU(2)'$. Its smallest singular value is considered to be (at least) of
the order of the scale $M$.
Factoring out this explicit mass scale, we parameterize the vacuum structure
responsible for the Yukawa matrices of the charged fermions as follows,
\begin{align}\label{eq:VEVs}
 \kappa_S\,\VEV{S} = s\,M\,
\quad\text{and}\quad 
 \kappa_T\,\VEV{T'} = \left(\begin{array}{cc} t' & 0 \\ 0 & -t' \end{array}\right)\,M \ ,
\end{align}
where $s$ and $t'$ are defined as dimensionless $3\times3$ matrices in flavour
space.

Turning to the neutrino sector, we adopt the double seesaw
mechanism~\cite{Mohapatra:1986bd} involving $\overline q_L$, $\Sigma_R$ and
$\overline \Theta_L$.\footnote{There exists another contribution to the
effective light neutrino masses which additionally involves the fields $\overline
\Sigma_L$,  $q_R$, $\overline \Xi_L$ and $\Xi_R$. Its structure will
automatically become manifest when integrating out the heavy degrees of
freedom in $\mathcal L_{\text{Yuk}}$.}
The PS-neutral fermions $\overline\Theta_L$ acquire  
a Majorana mass $\VEV{S_\nu}$ at a high scale $\Lambda_\nu$. Furthermore, a Dirac
mass term $\overline \Theta_L \VEV{\Phi'} \Sigma_R$ is generated at the scale
$\Lambda_\varphi \ll \Lambda_\nu$. With this hierarchy of scales, the first
stage of the double seesaw mechanism gives rise to a Majorana neutrino mass
for the SM neutral component $\Sigma_R^\nu$ within $\Sigma_R$.
In the second stage of the double seesaw, light Majorana neutrino masses for
the left-chiral neutrinos~$\overline q_L^\nu$ are induced via the Dirac
coupling of $\overline q_L^\nu$ and $\Sigma_R^\nu$. 
Analogous to Eq.~\eqref{eq:VEVs}, we parameterize the VEVs of the scalar
fields required for generating light Majorana neutrino masses as
\begin{align}
\VEV{S_\nu} =s_\nu\,\Lambda_\nu  \ , \qquad
\VEV{S'_\nu} =s'_\nu\,\Lambda_\nu  \ ,
\qquad\text{and}\qquad 
  \VEV{\Phi'} = \varphi'\, \Lambda_\varphi
 \,.
 \label{eq:GRVneutrinoVEVs}
\end{align}
Here we have included the VEV of $S'_\nu$ which will turn out to be crucial 
for the construction of the lepton sector as discussed in
Appendix~\ref{app:leptons}. The quantities $s_\nu$, $s'_\nu$ and $\varphi'$ are again
dimensionless tensor structures in flavour space. Note that $\Phi$ transforms
non-trivially under the electroweak symmetry. Similar to $T$, it must
therefore not receive a high scale VEV, and we will approximate $\VEV\Phi
\simeq0$ in the following.

In contrast to the original GRV model proposed in~\cite{Grinstein:2010ve}, in our setup the
flavour gauge bosons do not play an essential role for the low-energy phenomenology.
The reason for this is the inclusion of the lepton/neutrino sector
which requires the introduction of the flavons  $S_\nu,\,S^{\prime}_\nu$ and
$\Phi^{\prime}$ in addition to $S$ and $T^{\prime}$. While the latter
acquire VEVs reaching down to the scale $M$ (TeV regime), the former obtain
their VEVs at much larger scales, $\Lambda_\nu$ and~$\Lambda_\varphi$. As
these are related to the type I seesaw scale via $\Lambda_\varphi^2 /
\Lambda_\nu \sim M_{\text{seesaw}}$, it is reasonable to
assume a typical hierarchy of scales,
\be 
10^{10} \,\text{GeV} ~\lesssim ~
\Lambda_\varphi ~\ll~
\Lambda_\nu ~\lesssim~ M_{\text{Planck}} 
~\sim~ 10^{18} \,\text{GeV}  \ .\label{eq:lambdas}
\ee
Here, the lower bound is relatively flexible (due to the vague definition of
$M_{\text{seesaw}}$) and has been chosen rather conservatively. As a
consequence, the gauged flavour symmetry gets broken at a very large scale,
and the associated flavour gauge bosons become too heavy to be relevant for
current or future particle physics experiments. In principle, a subset of the
flavour gauge bosons could remain massless down to lower scales. However, if
the symmetric $3\times 3$ matrix $s_\nu$ has non-vanishing and non-degenerate
singular values, $SU(3)_I$ gets fully broken and all its
flavour gauge bosons acquire masses at the scale $\Lambda_\nu$. Similarly, all
flavour gauge bosons of $SU(3)_{II}$ become heavy due to $\VEV{S'_\nu}$. 

Concerning the masses of the gauge bosons associated with the PS symmetry, we
point out that we do not fully specify the breaking of $SU(4)$ or
$SU(2)'$. Yet, already the flavon~$\Phi'$, which we have introduced for the
purpose of generating light neutrino masses, breaks PS down to the
SM~\cite{Hartmann:2014fya,Slansky:1981yr}. 
Therefore, the masses of the non-SM gauge bosons, including the $W'^\pm$ 
and the $Z'$, are bounded from below by $\Lambda_\varphi$. Additional PS
breaking fields may lead to even larger masses.

\begin{figure}[t]
\centering
 \includegraphics[width=.8\textwidth]{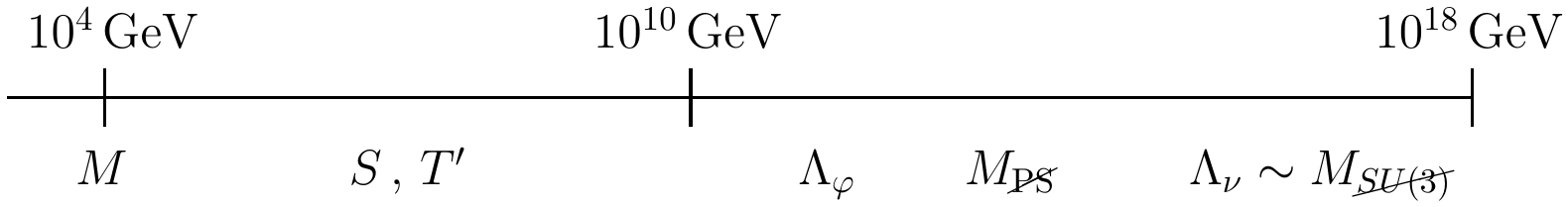}
\caption{Illustration of the hierarchy of scales introduced in the PS model
with gauged flavour symmetry. The numerical values are shown as an indicative
example only. $M_{\cancel{SU(3)}}$ denotes the scale of the flavour gauge
bosons, while $M_{\cancel{\text{PS}}}$ is the scale of the heavy PS gauge bosons.}
\label{fig:GRVscales}
\end{figure}

As discussed above, the various scales of the model are only partially
constrained. Their ordering is, however, determined as illustrated in
Fig.~\ref{fig:GRVscales}. In addition to the hierarchies of scales, the
singular values of $\VEV S$ and $\VEV{T'}$ feature internal hierarchies, with
the smallest reaching down to the scale $M$. The value of $M$ is bounded from
below by flavour precision observables as well as the non-observation of
new charged fermionic states. For the purpose of our numerical
parameter scan, we will assume values of a few TeV. Due to the strong
hierarchy of quark masses, the largest singular values of $\VEV S$ and
$\VEV{T'}$ take numerical values up to order $10^{10}\,\text{GeV}$. This is in fact
the reason for choosing the lower bound on~$\Lambda_\varphi$ as given in
Eq.~\eqref{eq:lambdas}. On the other hand, $\Lambda_\varphi$ is bounded from
above by the GUT scale $M_{\text{GUT}}$, as $\VEV{\Phi'}$ breaks PS down to
the~SM. Finally, the largest scale $\Lambda_\nu$ must lie between
$\Lambda_\varphi$ and the Planck scale $M_{\text{Planck}}$. In addition to
these scales, Fig.~\ref{fig:GRVscales} also shows the mass scale 
$M_{\cancel{SU(3)}}$ of the flavour gauge bosons as well as the mass scale
$M_{\cancel{\text{PS}}}$ of the heavy PS gauge bosons as discussed above.

Strictly speaking, the model is formulated at the GUT scale.  Due to the
presence of multiple scales, it is generally necessary to consider threshold
corrections at each of these scales. In what follows, we will, however,
neglect such effects as we do not aim for a comprehensive top-down description
of the model. We will therefore perform our numerical scan using flavour
parameters (e.g.~quark masses) which are fixed at the $Z$~boson mass scale.


\subsection{Approximate flavour structure of quarks and leptons}
\label{sec:approx}

In a first step to understand the structure of the low-energy effective
Lagrangian derived from Eq.~\eqref{eq:lagrangian}, we formally integrate out
the heavy degrees of freedom. The result will provide reasonably simple
expressions for the approximate flavour structure of quarks and leptons. We
begin our discussion with the quark sector, where all relevant terms are
included in $\mathcal L_{\text{Yuk}}^q$ of Eq.~\eqref{eq:fermionlagrangian}. 
Assuming for the moment all VEVs of $S$ and $T'$ to be significantly larger than the
intrinsic mass scale $M$, i.e.\ $s \sim t' \gg 1$, we can integrate out the heavy
fermions $\Sigma_{L,R}$ and $\Xi_{L,R}$.\footnote{To be precise, considering
the quark sector, we only have to integrate out the colour (anti-)triplets contained
in the (anti-)fundamental representations of $SU(4)$.} Defining the 
$3\times 3$ matrices
\be
 t_u~ \equiv ~ s+t' \  , \qquad  t_d~ \equiv ~s-t' \ , 
\ee
the resulting effective Yukawa potential takes the form
\begin{align}\label{eq:effpotential}
 \mathcal{L}_\text{Yuk}^{q,\text{eff}} =
     &- \lambda\;h_u^0 \;\overline{q}_L^u\,
    \left[ t_u^{-1} + s^{-1}\right]
       \;q_R^u\; +h.c.\cr  
     &-\lambda\;h_d^0\;\overline{q}_L^d\, 
     \left[ t_d^{-1}+ s^{-1}\right]
     \;q_R^d \;+h.c. \ ,
\end{align}
where the $SU(2)$ and $SU(2)'$ doublets have been explicitly decomposed into their up
and down components. The effective couplings correspond to the standard
quark Yukawa matrices in a two-Higgs-doublet model, and can be directly read off as
\begin{equation} 
  Y_{u,d} ~\simeq ~ -\,\lambda\;\left[t_{u,d}^{-1} + s^{-1}\right]  \ .
\label{eq:Yukrelations}
\end{equation}
It is worth mentioning that this relation between the flavon VEVs and the SM Yukawa
matrices is neither a linear one as in Minimal-Flavour Violation (MFV)~\cite{Chivukula:1987py,Hall:1990ac,Buras:2000dm,D'Ambrosio:2002ex,Cirigliano:2005ck}
nor a simple inverse one as in~\cite{Grinstein:2010ve}, where the difference
with the latter can be traced back to the imposed $Z_2$ symmetry. Yet,
similarly to the original GRV case, the expressions in
Eq.~\eqref{eq:Yukrelations} show that the observed hierarchy of the Yukawa 
matrices is turned into an inverse hierarchy of the $3\times 3$ matrices $s$ and $t'$. 
Eq.~\eqref{eq:Yukrelations} can now be used to fit the 19 independent
parameters of $s$ and $t'$ (six singular values, six mixing angles and seven
phases) to the quark masses and the CKM matrix $V_{\text{CKM}}$. In contrast
to the SM, the mixing of the right-chiral quarks (parameterized by
$U'_{\text{CKM}}$ in the following) is physical in our PS model and has to be kept as a
consequence. As the relation between $Y_{u,d}$ and $s,t'$ is given by a
coupled matrix equation, it is a non-trivial task to invert 
Eq.~\eqref{eq:Yukrelations}. We describe the method used to obtain numerical 
results for $s$ and $t'$ in Appendix~\ref{sec:seesaw}.

An important feature, already arising at this stage, is the existence of 
multiple solutions when inverting Eq.~\eqref{eq:Yukrelations}. This can be easily understood in
the case of one generation, where the resulting two solutions can be expressed
analytically. Explicitly, we find 
\begin{subequations}\label{eq:GRVInversion}
\begin{align}
  s &= \frac{-8\lambda}{3y_u + 3 y_d \mp 3\sqrt{\left(y_u-y_d\right)^2 + \frac{4}{9}\, y_uy_d}} \, ,\\
  t_u &= \frac{-8\lambda}{5y_u - 3 y_d \pm 3\sqrt{\left(y_u-y_d\right)^2 +    \frac{4}{9}\, y_uy_d}} \, ,  
\end{align}
\end{subequations}
with $y_{u,d}$ denoting the one-generation up-type and down-type Yukawa
couplings. Using these expressions it is straightforward to determine $t'$ as well as
$t_d$. Extending this result to three generations gives rise to a total of eight
solutions for $s$ and $t'$. We emphasize that all of these generate the same
approximate Yukawa matrices via Eq.~\eqref{eq:Yukrelations}. However, as we
will see in Section~\ref{sec:GRVQuarkPheno}, different solutions may generate
different phenomenological predictions beyond the SM.

Turning to the lepton sector, we first remark that the structure of
the charged lepton Yukawa matrix is identical to the one of the down-type
quarks. This is a common feature to the simplest Pati-Salam models and a
reasonably good first approximation. Possible modifications to the charged
lepton sector are briefly sketched in Appendix~\ref{app:alt_chargedleptons}.
In the neutrino sector, the Lagrangian involving the heavy
fermions $\Theta_{L,R}$ generates a Majorana mass term for the neutrino
component $\Sigma_R^\nu$ within the PS multiplet $\Sigma_R$. Integrating out 
$\Theta_{L,R}$ in $\mathcal L_{\text{Yuk}}^\nu$ of
Eq.~\eqref{eq:fermionlagrangian-nu} yields
\be\label{eq:effnuL}
\mathcal L_{\text{Yuk}}^{\nu,\text{eff}} ~\sim~ 
-\,\frac{(\varphi_\alpha' \Lambda_\varphi)^2}{2\, \Lambda_\nu} 
 ~ \Sigma^\nu_R \, s_\nu^{-1} \, \Sigma^\nu_R 
\ .
\ee
We remark that $s_\nu$ is a symmetric $3\times 3$ matrix in $SU(3)_I$
flavour space, while $\varphi'_\alpha \Lambda_\varphi$ denotes the VEV of
$\Phi'$ which points in the direction of the SM neutral component of the PS
representation $(\rb4,\re1,\re2)$.  Being an octet of $SU(3)_{II}$,
$\varphi'_\alpha$ with $\alpha=1,...8$ does not affect the flavour structure
of the neutrino sector. Combining this Majorana mass term
for $\Sigma_R^\nu$ with the couplings of the electro\-magnetically neutral
fermions in $\mathcal L^q_{\text{Yuk}}$ of Eq.~\eqref{eq:fermionlagrangian}
generates Majorana masses for the left-chiral neutrinos. The resulting
structure of the neutrino mass matrix can be obtained by formally integrating
out the heavy neutral fermions $\Sigma_{R}^\nu$ and $\Xi_{L,R}^\nu$ as well as
$\Sigma_{L}^\nu$ and $q_R^\nu$. It is given by
\be \label{eq:effneutrinomass}
  m_\nu^\text{eff} ~\sim ~\frac{\Lambda_\nu \,v_u^2}{2\,(\varphi_\alpha' \Lambda_\varphi)^2} \left(\id +
  s^{-1}t_u\right)\,s_\nu\,\left(\id + s^{-1}t_u\right)^T \ ,
\ee
where $v_u/\sqrt{2}$ denotes the VEV of the neutral Higgs $h_u^0$, cf. Appendix~\ref{app:leptons}.
 It is important to
note that the flavour structure of the light neutrinos is decoupled from the
quark sector. In particular, the distinct hierarchical structure of $s$ and
$t_u$ is approximately cancelled in the combination~$s^{-1} t_u$, such that
the structure of the effective neutrino mass matrix is mainly determined by
$s_\nu$. As a consequence, the PMNS mixing of the lepton sector can be
significantly different from the CKM mixing of the quarks. 
Assuming the dimensionless parameters $s^{-1}t_u$, $s_\nu$ and $\varphi'_\alpha$ 
to be of order one, we find that $\Lambda_\varphi \sim 10^{16}\GeV\sim M_\text{GUT}$
and $\Lambda_\nu \sim 10^{18}\GeV\sim M_\text{Planck}$  gives rise to realistic
neutrino masses of the order $0.1\,\eV$. As such, the model provides sufficient
freedom to construct a phenomenologically viable lepton sector.
More detailed studies are left for future work.




\section{Quark flavour sector}
\label{sec:GRVQuarkFlavour}
\cleqn

The effective quark flavour structure derived in Section~\ref{sec:approx} is
only an approximation which relies on the simplifying assumption that $s,t' \gg 1$.
Clearly, such a relation is not satisfied for the Yukawa couplings of the third
generation of quarks which should be of order one for the top quark (and,
depending on $\tan \beta =v_u/v_d$, also for the bottom quark). 
Taking this fact into account, it is necessary to diagonalize the full
$9\times 9$ mass matrices of the SM quarks and their heavy partners. The
corresponding transformations are then applied to the gauge-kinetic sector in
order to extract the characteristic features of the effective low-energy quark
flavour sector. We also comment on the coupling of the Higgs field to the SM quarks,
which is no longer fully aligned with the diagonal quark mass matrices.


\subsection{\label{subsec:sequence}Diagonalizing the quark mass matrices}

In this subsection, we re-examine the Yukawa Lagrangian of
Eq.~\eqref{eq:fermionlagrangian}. Assigning VEVs to the flavour symmetry
breaking fields $S$ and $T'$ as well as the $SU(2)\times SU(2)'$
bi-doublet~$H$, Eq.~\eqref{eq:fermionlagrangian} gives rise to 
\begin{subequations}
\begin{align}
 \mathcal{L}^q_{\text{mass}} = 
 & \:\:\phantom{+} \tfrac{1}{\sqrt{2}}\, \lambda\,v_u\;\overline{q}^u_L\,\Sigma_R^u +  M\;\overline{\Sigma}^u_L\,t_u\,\Sigma^u_R + M\;\overline{\Sigma}^u_L\,q^u_R  + h.c.\\
 &+\tfrac{1}{\sqrt{2}}\, \lambda\,v_u\;\overline{\Xi}^u_L\,q_R^u +M\;\overline{\Xi}^u_L\,s\,\Xi^u_R + M\;\overline{q}^u_L\,\Xi^u_R   + h.c.\\
& + (u \leftrightarrow d) \ . \notag
\end{align}
\end{subequations}
Collecting the left-chiral (right-chiral) fermions in $\Psi_L$ ($\Psi_R$),
these bilinear mass terms can be compactly written as 
${\overline \Psi^u_L} \mathcal M^u {\Psi^u_R}$ and 
${\overline \Psi^d_L} \mathcal M^d {\Psi^d_R}$. 
For definiteness, we focus on the up-type quark sector in the rest of this
subsection. Analogous results arise for the down-type quark sector, and the
corresponding expressions are obtained by simply replacing the index $u$ by $d$.
Defining
\be \label{eq:defPSI}
{\overline \Psi^u_L} ~\equiv ~(\overline q_L^u , \overline \Sigma_L^u , 
\overline \Xi_L^u) \ , ~ \qquad
{\Psi^u_R} ~\equiv ~( q_R^u ,  \Sigma_R^u ,  \Xi_R^u) \ ,
\ee
the $9\times 9$ mass matrix takes the form 
\be\label{eq:Mu}
\mathcal M^u ~=~ 
\begin{pmatrix} 
0 & \id \,  \lambda \epsilon_u & \id \\
\id  &  t_u  & 0  \\
\id\,  \lambda  \epsilon_u  & 0 & s
\end{pmatrix} M \ ,
\ee 
where each entry corresponds itself to a $3\times 3$ matrix in generation space. 
The quantity
\be
\epsilon_{u}\equiv \frac{v_{u}}{\sqrt{2}\,M}\ll 1 \ 
\ee
can be treated as a small expansion parameter.
In the following, we describe the sequence of basis transformations which
diagonalizes $\mathcal M^u$ of Eq.~\eqref{eq:Mu} in powers of $\epsilon_u$ up to
quadratic order. More technical details of this diagonalization are relegated
to Appendix~\ref{app:sequence}.

In the first step, we make use of the $SU(3)_I\times SU(3)_{II}$ flavour
symmetry to diagonalize the $3\times 3$ matrix $s$ in Eq.~\eqref{eq:Mu}.
This corresponds to a choice of basis which can always be made without loss of
generality. We can thus simply replace $s$ by $\hat s$, where here and in the following
the hat denotes a diagonal matrix. Next, we apply a bi-unitary basis transformation to diagonalize the
submatrix $t_u$, i.e.\
\be\label{eq:diag-tu}
\hat t_{u} ~=~ V_u  \, t_{u} \, U_u^\dagger \ .
\ee
Having exhausted the flavour symmetry by choosing a diagonal $\hat s$, the
unitary $3\times 3$ matrices $V_u$ and $U_u$ will reappear elsewhere in the full
mass matrix. As discussed in Appendix~\ref{app:sequence}, they can however be
shifted to the $\epsilon_u$-suppressed blocks. In the limiting case of
$\epsilon_u=0$, the resulting mass matrix takes the form of Eq.~\eqref{eq:Mu}
with $s$ and $t_u$ replaced by $\hat s$ and $\hat t_u$. In this limit, the
three generations decouple from one another and the $9\times 9$ matrix 
decomposes into three $3\times 3$ blocks, one for each generation~$i$. 
Introducing only two mixing angles for each generation (whose values depend
on $\hat s$ and~$\hat t_u$), it is straightforward
to diagonalize these blocks exactly. In the third step, we therefore apply 
such a transformation to the full mass matrix including the $\epsilon_u$-suppressed 
blocks. This gives rise to a mass matrix of the form
\begin{align}
 \mathcal{M}^u ~\rightarrow~ \left(\begin{array}{ccc} a_u\epsilon_u & b_u\epsilon_u & 0 \\ 0 & 
\hat e_u & 0 \\ c_u\epsilon_u & d_u \epsilon_u & \hat f \end{array}\right) M
 \ ,
\label{eq:Mu3}
\end{align}
where the $3\times3$ matrices $a_u$, $b_u$, $c_u$, $d_u$ $\hat e_u$ and $\hat f$ 
depend on $V_u$, $U_u$ as well as the above mentioned mixing. The explicit
expressions are given in Appendix~\ref{app:sequence}. Note
that the mass matrix in Eq.~\eqref{eq:Mu3} is diagonal at zeroth order in the
expansion parameter. Block-diagonalization to second order in $\epsilon_u$ requires an
intricate basis transformation which can be found in
Appendix~\ref{app:sequence}. Yet, the resulting mass matrix simply reads 
\be\label{eq:Mu4}
\mathcal M^u ~\rightarrow~ 
\begin{pmatrix} 
a_u \epsilon_u & 0 & 0 \\
0& \hat e_u + \mathcal O(\epsilon_u^2)& 0 \\
0&0&\hat f + \mathcal O(\epsilon_u^2)
\end{pmatrix} M ~+~  \mathcal O (\epsilon_u^3) \,M \ .
\ee
The remaining transformation needed to render this matrix 
completely diagonal (up to order~$\epsilon_u^2$) involves only rotations within
the three non-vanishing $3\times 3$  blocks. These do not mix the light quarks
with the heavy partners. 
Due to our lack of experimental data on any heavy fermions, it is safe to ignore
the corresponding $3\times 3$ rotations which would take the general form $\id
+ \mathcal O(\epsilon_u^2)$. For the three light quarks on the other hand, we
define the bi-unitary transformation which diagonalizes $a_u$, 
\be
\hat Y_u ~=~ \mathcal V_u \, a_u\, \mathcal U_u^\dagger \ .
\label{eq:Yhat}
\ee
The complete sequence of transformations diagonalizes the $9\times 9$ quark mass
matrix of Eq.~\eqref{eq:Mu} up to second order in $\epsilon_u$. It therefore
constitutes the change from the original flavour basis $\Psi^{u,d}_{L,R}$ to
the approximate mass basis $\Psi'^{u,d}_{L,R}$. The individual steps of this
change of basis are explicitly given in Appendix~\ref{app:sequence}. In the
next subsection, we will apply these transformations to the gauge-kinetic sector.


\subsection{Gauge-kinetic terms}
\label{sec:gaugekin}

As discussed in Section~\ref{sec:SBandMS}, the flavour gauge bosons are far
beyond the reach of current or future particle physics experiments. They are
therefore phenomenologically irrelevant and we do not consider them any further.
Concerning the 15 gauge bosons associated with the $SU(4)$ factor of the PS
symmetry we note that 6 become extremely massive and irrelevant when $SU(4) \rightarrow
SU(3)_c \times U(1)_{B-L}$. Moreover, the 8 gluons of $SU(3)_c$ are flavour
blind and can be ignored as well. The remaining $U(1)_{B-L}$ boson will mix with
the neutral gauge bosons of $SU(2)\times SU(2)'$ and has to be kept. Being
mainly interested in the flavour structure, it is therefore sufficient to
focus on the gauge-kinetic terms corresponding to the 
$SU(2)\times SU(2)'\times U(1)_{B-L} $ part of the PS symmetry. In order to 
be able to apply the sequence of basis transformations described above (and in
Appendix~\ref{app:sequence}), we express the relevant gauge-kinetic terms by
means of the left- and right-chiral vectors $\Psi_L$ and $\Psi_R$ defined in
Eq.~\eqref{eq:defPSI}. In this context, it is important to realize that $\Psi_L$ does not solely
contain doublets of $SU(2)$ but also doublets of $SU(2)'$. Likewise, $\Psi_R$
contains doublets of both $SU(2)'$  and $SU(2)$. Taking this fact into
account, the relevant gauge-kinetic terms read
\begin{align}
\label{eq:couplingtoWp}
\mathcal{L}_\text{kin} ~ \supset &\phantom{+~\;}
{\overline \Psi^{}_L} \left( g  \vec{\slashed{W}} \vec{\tau} \right)
{\mathcal K_L}    {\Psi^{}_L} 
&&\hspace{-10mm}+~
{\overline \Psi^{}_R} \left( g  \vec{\slashed{W}} \vec{\tau} \right)
{\mathcal K_R}    {\Psi^{}_R} \nonumber\\
&+~
{\overline \Psi^{}_L} \left( g  \vec{\slashed{W'}} \vec{\tau} \right)
{\mathcal K'_L}    {\Psi^{}_L} 
&&\hspace{-10mm}+~
{\overline \Psi^{}_R} \left( g  \vec{\slashed{W'}} \vec{\tau} \right)
{\mathcal K'_R}    {\Psi^{}_R} \nonumber \\
&+~
{\overline \Psi^{}_L} \left(\tfrac12\, g_{B-L} Q_{B-L} \vec{\slashed{B}}_{B-L} \right)  {\Psi^{}_L} 
&&\hspace{-10mm}+~
{\overline \Psi^{}_R} \left(\tfrac12\, g_{B-L} Q_{B-L} \vec{\slashed{B}}_{B-L} \right)  {\Psi^{}_R} \ ,
\end{align}
where $Q_{B-L}$ is the difference of baryon and lepton number, and the
$\mathcal K$ matrices
\begin{subequations}
\begin{align}
{\mathcal K_L} &= 
\begin{pmatrix} \id&0&0 \\ 0&0&0 \\0&0&\id \end{pmatrix} \ , 
&{\mathcal K_R} &= 
\begin{pmatrix} 0&0&0 \\ 0&0&0 \\0&0&\id \end{pmatrix} \ , \\
{\mathcal K'_L} &= 
\begin{pmatrix} 0&0&0 \\ 0&\id &0 \\0&0&0 \end{pmatrix} \ , 
&{\mathcal K'_R} &= 
\begin{pmatrix} \id&0&0 \\ 0&\id&0 \\0&0&0 \end{pmatrix} \, ,
\end{align}
\end{subequations}
encode the non-trivial $SU(2)\times SU(2)'$ transformation properties of
$\Psi_L$ and $\Psi_R$. Here we emphasize that the $SU(2)$ gauge bosons couple to
both left- and right-chiral fermions. Likewise, the $SU(2)'$ gauge bosons
couple to fermions of both chiralities.

While breaking $SU(2)'\times U(1)_{B-L}\rightarrow U(1)_Y$ the neutral bosons
$W'^{(3)}$ and $B_{B-L}$ mix, resulting in a massless $B_Y$ and a massive $Z'$
boson. Analogous to the electroweak symmetry breaking in the SM, this mixing
is described by $s'_W$ and $c'_W$ which satisfy 
$g\,s_W' = g_{B-L}\,c_W' \equiv  g_Y$.
The charged $SU(2)'$ gauge bosons $W'^\pm \equiv (W'^1\mp i \, W'^2)/\sqrt{2}$
become massive from the $SU(2)'$ breaking as well. 
As this breaking arises at an extremely high scale,
cf. Figure~\ref{fig:GRVscales}, the ${W'}^\pm$ and $Z'$ gauge bosons are
irrelevant for low-energy flavour effects. 
In a second step, the breaking $SU(2)\times U(1)_{Y}\rightarrow
U(1)_{\text{em}}$ at the electroweak scale induces a mixing of the neutral
gauge bosons $W^3$  and $B_Y$ which is parameterized by $s_W$ and $c_W$
satisfying $g\,s_W = g_Y \,c_W \equiv  e$.
Using these relations and the two-step mixing
\begin{subequations}
\bea
\slashed B_{B-L} &=& c'_W \,(c_W \slashed A - s_W \slashed Z ) ~-~
s'_W\,\slashed Z'  \ ,\\
{\slashed W}^{\prime\,3}\, &=& s'_W \,(c_W \slashed A - s_W \slashed Z ) ~+~
c'_W\,\slashed Z'  \ ,\\
\slashed W^3~ &=& \hspace{7.5mm}s_W \slashed A + c_W \slashed Z \ ,
\eea
\end{subequations}
we can rewrite the gauge-kinetic Lagrangian, keeping only the terms which are
relevant for low-energy flavour effects,
\begin{align}
  \mathcal{L}_\text{kin} ~\supset &
\phantom{+~\ }
 {\overline \Psi^{}_L} \frac{g}{\sqrt{2}}  ~ \mathcal K_L
 \left(\tau^1 \pm i\,\tau^2\right) 
    \slashed{W}^\pm  {\Psi^{}_L} \nonumber\\
&+~
{\overline \Psi^{}_L} \frac{g}{c_W} \biggl(
 \left(c_W^2\, {\mathcal K_L} - s_W^2 \,{\mathcal K'_L}\right)  \tau^3
-  \tfrac12s_W^2 Q_{B-L} \biggr)\slashed{Z} {\Psi^{}_L} \nonumber\\
&+~
{\overline \Psi^{}_L} \,e \biggl(  \left({\mathcal K_L} +
{\mathcal K'_L}\right) \tau^3 +  \tfrac12Q_{B-L} \biggr)\slashed{A}
{\Psi^{}_L} \nonumber\\
&+~
(L \leftrightarrow R) \ .
\end{align}
In order to simplify the notation, it is convenient to decompose the isospin
doublets in the charged current interactions explicitly into their
components. Using the relations 
$\mathcal K_L+\mathcal K'_L = \mathcal K_R+\mathcal K'_R = \id$ 
as well as\footnote{The relation between the electric charge $Q_e$ and
  $Q_{B-L}$ takes this simple form as the fermions of the model transform
  solely under one of the two $SU(2)$ gauge factors. Therefore, $\tau^3$ is
  understood to act on either $SU(2)$ or $SU(2)'$, depending on the specific particle.}
$Q_e = \tfrac{1}{2}\,Q_{{B-L}} + \tau^3$, we obtain  
\begin{align}
\label{eq:GRVbrokenkinlag}
\mathcal{L}_\text{kin} ~ \supset
&\phantom{+\;}
\ \,\frac{g}{\sqrt{2}}\;\,{\overline \Psi^{u}_L} 
\,{ \mathcal K_L}  \,
\slashed{W}^+ {\Psi^{d}_L} + \hc \nonumber\\
&+
\  \frac{g}{\sqrt{2}}\;\,{\overline \Psi^{u}_R} 
\,{\mathcal K_R} \,
\slashed{W}^+ {\Psi^{d}_R} + \hc \nonumber\\
&+ \
\frac{g}{c_W}\;\,{\overline \Psi^{}_L} \biggl( \left(\tau^3\, - s_W^2 Q_e\right)\id - {\mathcal K'_L}\,\tau^3 \biggr)\slashed{Z} {\Psi^{}_L} \nonumber\\
&+\
\frac{g}{c_W}\;\,{\overline \Psi^{}_R} \biggl( ~~-\,s_W^2\;Q_e \id 
\hspace{6.3mm}+ {\mathcal K_R}  \, \tau^3\biggr)\slashed{Z}{\Psi^{}_R} \nonumber\\
&+
\ e \left( {\overline \Psi^{}_L}\,Q_{e}\, \slashed{A} {\Psi^{}_L} +~ {\overline \Psi^{}_R} \,Q_{e}\,\slashed{A} {\Psi^{}_R}\right) \,.
\end{align}
When applying the sequence of basis transformations discussed above, it is
important to take into account the isospin structure of the charged and
neutral couplings. As a result, the flavour structure of all terms involving
the $\mathcal K$ matrices is modified in a non-trivial way, while the
form of the remaining terms is left unchanged. For instance, the last line of
Eq.~\eqref{eq:GRVbrokenkinlag} shows that the coupling of the photon is always
diagonal and proportional to the corresponding electric charge. The explicit
expressions of the full $\mathcal K$ matrices in the approximate mass basis are
provided in Appendix~\ref{app:sequence}. They will be important when studying
the gauge interactions of the light quarks with their heavy partners such as
e.g.\ the coupling~$\overline b \, t' \, W^-$. Being mainly interested in the
phenomenological flavour effects involving only the three generations of light
quarks $q'_{L,R}$ (where the prime denotes the approximate mass basis), we
focus our attention on the upper left $3\times 3$ block. 
Separating the couplings which are already present in the SM (such as
$V_\text{CKM}$) from terms which are characteristic to our setup 
(i.e.\ $\Delta V_\text{CKM}$,
$U_\text{CKM}$, 
$\Delta g_{Z\overline{q}^{}_Lq^{}_L}$ and 
$\Delta g_{Z\overline{q}^{}_Rq^{}_R}$), the terms of the gauge-kinetic
Lagrangian take the form
\begin{align}\label{eq:L_kingauge}
\mathcal{L}_\text{kin} ~ \supset 
&\phantom{\:\:+~}
 \frac{g}{\sqrt{2}}\;{\overline q'^{u}_L} \left(V_\text{CKM} - \Delta V_\text{CKM} \right)
 \slashed{W}^+{q'^{d}_L} + \hc \nonumber\\ 
& +~
 \frac{g}{\sqrt{2}}\;{\overline q'^{u}_R}\; U_\text{CKM}\, \slashed{W}^+ {q'^{d}_R}  + \hc \nonumber\\
& +~
\frac{g}{c_W}\, {\overline q'^{}_L}\biggl( \left(\tau^3\, - s_W^2
Q_e\right)\id - \Delta g_{Z\overline{q}^{}_Lq^{}_L}\, \tau^3 \biggr)\slashed{Z}\, {q'^{}_L} \nonumber\\
& +~
\frac{g}{c_W}\, {\overline q'^{}_R} \biggl(~~ -\,s_W^2\;Q_e
\id \hspace{6.3mm}+\Delta g_{Z\overline{q}^{}_Rq^{}_R} \,
\tau^3\biggr)\slashed{Z}\, {q'^{}_R}\ .
\end{align}
We emphasize that the two matrices $\Delta g_{Z\overline{q}^{}_Lq^{}_L}$ and 
$\Delta g_{Z\overline{q}^{}_Rq^{}_R}$ will generally be different for the two
isospin components. Mindful of this subtlety, we
however suppress the corresponding isospin indices for the sake of notational
simplicity. Comparing the explicit expressions of Appendix~\ref{app:sequence},
in particular Eq.~\eqref{eq:gaugekinflavour}, with Eq.~\eqref{eq:L_kingauge}
reveals that the low-energy flavour effects are parameterized as follows,
\begin{subequations}
\label{eq:GRVflavoureffects}
\begin{align}
V_\text{CKM} &=~ \mathcal{V}_u\,\mathcal{V}_d^\dagger \ , \\
\Delta V_\text{CKM} &=~ \tfrac{1}{2} \,  \mathcal{V}_u \left(b_u\,\hat{e}_u^{-2}\,b_u^\dagger\,\epsilon_u^2  + b_d\,\hat{e}_d^{-2}\,b_d^\dagger\,\epsilon_d^2 \right)  \mathcal{V}_d^\dagger ~+~ \mathcal{O}\left(\epsilon^3_{u,d}\right)\, , \\[.5ex]
\label{eq:GRVCKMp}
U_\text{CKM} &= \mathcal U_u \,c_u^\dagger \hat{f}^{-2}\,c_d\, {\mathcal
  U_d}^\dagger \,\epsilon_u \epsilon_d + \mathcal{O}\left(\epsilon^3_{u,d}\right) \, ,\\[.5ex]
\Delta g_{Z\overline{q}^{}_Lq^{}_L}  &=\mathcal V\, b\,\hat{e}^{-2}\,b^\dagger  \,{\mathcal V}^\dagger \,\epsilon^2 + \mathcal{O}\left(\epsilon^3\right) \,,\\
\label{eq:GRVZqr}
\Delta g_{Z\overline{q}^{}_Rq^{}_R}  &= \mathcal U\, c^\dagger \hat{f}^{-2}\,c\,{\mathcal U}^\dagger \,\epsilon^2 + \mathcal{O}\left(\epsilon^3\right) \,.
\end{align}
\end{subequations}
We point out that the effective CKM matrix $V^{\text{eff}}_{\text{CKM}}$ is no 
longer unitary as it receives a correction $\Delta V_\text{CKM}$. Like all the
other beyond the SM flavour effects, this correction is of second order in the expansion
parameters $\epsilon_{u,d}$.


\subsection{Non-standard Higgs couplings}
\label{sec:GRVHiggscoupling}

As can be seen from the Lagrangian of Eq.~\eqref{eq:fermionlagrangian},  there
exists no direct coupling of the Higgs boson to a pair of SM quarks. Similar
to the quark masses, such an interaction is generated effectively through
couplings to the heavy quark partners. For this reason, the Higgs-quark-quark
coupling $g_{h\overline{q}q}$ is no longer exactly diagonal in generation space nor
proportional to the quark masses. The effective expression for
$g_{h\overline{q}q}$ can be deduced by formally integrating out the heavy
fermions or, alternatively, by explicitly applying the sequence of basis
transformations defined above to the Higgs couplings. As a result, we find
that the corrections to the quark masses as well as the Higgs-quark-quark
coupling are both of order $\epsilon^2_{u,d}$, however, they differ from one another
by a combinatorial factor of three.

\begin{figure}[t]
  \centering
  \includegraphics[width=\textwidth]{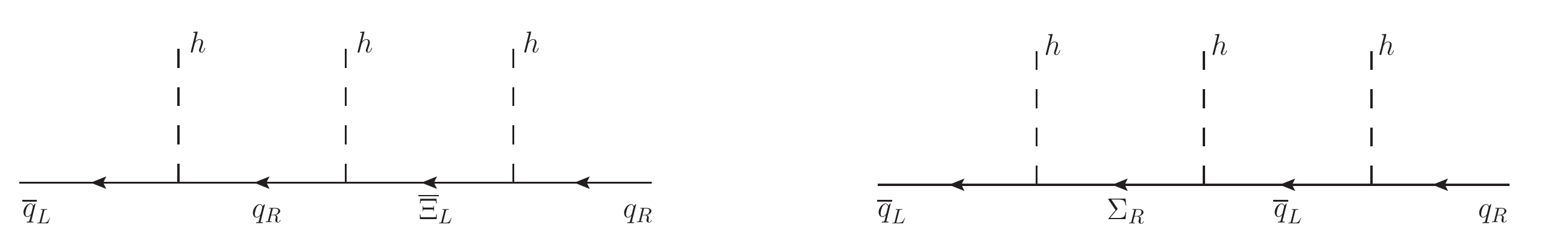}
  \caption{Leading order corrections to the quark masses and the
    Higgs-quark-quark coupling $g_{h\overline{q}q}$.}
  \label{fig:GRVHiggs}
\end{figure}

This effect can be understood by explicitly studying the leading order
corrections starting from the basis of Eq.~\eqref{eq:Mu3}. We stress that this
basis does not make use of any approximation in terms of a truncated $\epsilon_{u,d}$
expansion. Reinserting the original Higgs field in Eq.~\eqref{eq:Mu3}, the
corresponding diagrams, depicted in Figure~\ref{fig:GRVHiggs}, lead to an
effective coupling of three Higgs bosons to a pair of SM quarks. 
They  provide a correction to the quark masses if all Higgs fields acquire their VEV.
Alternatively, if only two of them are set to their VEV, the diagrams of
Figure~\ref{fig:GRVHiggs} contribute to the Higgs-quark-quark
coupling~$g_{h\overline{q}q}$. With three possible ways of choosing two Higgs
VEVs (and one Higgs field), the latter correction picks up the combinatorial factor of
three. Such an effect also occurs in the SM when considering general effective
dimension six operators~\cite{Buchmuller:1985jz}. 

As the effect arise identically in the up-type as well as the down-type quark
sector, we suppress the subscripts $u$ and $d$ in our resulting expression
which are valid up to order~$\epsilon^2_{u,d}$. In the approximate mass basis, where
the three light generations of quarks are denoted by $q'_{L,R}$, the effective
Yukawa coupling takes the form 
\begin{align}
  \label{eq:GRVYukeps2}
  Y ~\approx ~ \hat Y ~-~ \frac{\epsilon^2}{2} \, 
\Bigl[\,
  \mathcal{V}\,  b\,\hat{e}^{-2}_{}\,b^\dagger\, \mathcal V^\dagger \cdot \hat Y 
+  \hat Y \cdot\, \mathcal U   \,c^\dagger\,\hat f^{-2}_{}\,c \, \mathcal{U}^\dagger 
\,\Bigr] \, ,
\end{align}
while the Higgs-quark-quark coupling is given by
\begin{align}
  \label{eq:GRVhnlo}
  g_{h\overline{q}q} ~\approx ~ 
\hat Y ~-~ \frac{3\, \epsilon^2}{2} \, 
\Bigl[\,
  \mathcal{V}\,  b\,\hat{e}^{-2}_{}\,b^\dagger\, \mathcal V^\dagger \cdot \hat Y 
+  \hat Y \cdot\, \mathcal U   \,c^\dagger\,\hat f^{-2}_{}\,c \, \mathcal{U}^\dagger 
\,\Bigr] \,.
\end{align}
In these expressions, $\hat Y$ is the exactly diagonalized version of the matrix $a$,
using the unitary transformations $\mathcal V$ and $\mathcal  U$,
cf. Eq.~\eqref{eq:Yhat}. We observe that the Yukawa matrix in
Eq.~\eqref{eq:GRVYukeps2} is only diagonal up to first order in
$\epsilon_{u,d}$. Diagonalizing $Y$ to second (or in fact arbitrary) order 
can be achieved by small modifications of $\mathcal V$ and $\mathcal U$.
Applying the same basis transformation  to~$g_{h\overline{q}q}$ of
Eq.~\eqref{eq:GRVhnlo} will generally leave the Higgs-quark-quark coupling
non-diagonal at order~$\epsilon^2_{u,d}$.  Moreover, comparing
Eqs.~\eqref{eq:GRVYukeps2} and \eqref{eq:GRVhnlo}, is it possible to show that
the diagonal couplings of $g_{h\overline{q}q}$ are reduced compared to their
SM values.  Numerically, we explicitly find such an effect in our scan over
model parameters, where we exemplarily show the (normalized) Higgs-top-top coupling in
Figure~\ref{fig:httvM}.




\section{Scanning the parameter space}
\label{sec:scan}
\cleqn

In the previous section, we have calculated analytical expressions for the
Yukawa matrices and deduced flavour parameters of the model in terms of the
underlying VEVs $\VEV{S}$ and $\VEV{T'}$. However, we are not able to solve
the derived relations, such as e.g.\ Eq.~\eqref{eq:Yhat}, analytically for $s$
and $t'$. In other words, we cannot parameterize the flavour signatures of the
high-energy model using the physically known low-energy flavour parameters, i.e.\ the quark masses
$m^{u,d}_i$ and the CKM mixing $V_{\text{CKM}}$.

On the other hand, it is possible to invert the approximate formula of
Eq.~\eqref{eq:Yukrelations} numerically. Doing so, we encounter the technical
complication of multiple solutions, as already mentioned at the end of
Section~\ref{sec:approx}. Moreover, starting from a left-right symmetric
theory the right-chiral up-type and down-type quark sector cannot be rotated
independently. This enforces the introduction of the mixing matrix
$U'_\text{CKM}$ which parameterizes the coupling of the ${W'}^\pm$ bosons to
the right-chiral quarks and contains six physical phases. As both the mixing
matrix $U'_\text{CKM}$ as well as the appropriate choice of a solution are
unknown, we choose to scan over these degrees of freedom. In this section, we
describe the technical details of our procedure. In principle a scan of the
parameter space should be performed over all parameters of the model, namely $s$, $t'$,
$\lambda$, $\tan\beta=v_u/v_d$ and $M$. However, as $s$ and $t'$ feature large
hierarchies a direct scan over these is not reasonable. Thus we use the
alternative ansatz and scan over so-called ``adapted flavour parameters''
which we define in the following. We point out that these are deduced from the
parameters of the SM Yukawa couplings.


\subsection{Adapted flavour parameters}

We start our considerations with the quark masses and mixings as fitted
in the SM framework~\cite{Agashe:2014kda}. In order to take into
account the physical but 
experimentally unknown mixing of the right-chiral quarks in the $SU(2)'$
charged current, we also add mixing angles and phases for the
matrix~$U'_\text{CKM}$. With this input we calculate an explicit numerical realization
for the $3\times 3$ matrices $Y_u$ and $Y_d$ in an arbitrary basis.
For a given such pair it is possible to invert Eq.~\eqref{eq:Yukrelations},
thereby deriving $s$ and $t'$ numerically, see Appendix~\ref{sec:seesaw}. 
As explained in Section~\ref{sec:approx}, these are however only
approximations, and we expect sizeable corrections for the third generations. 
Nevertheless they provide a well-motivated starting point for the exact 
computation. 

Inserting the so-derived numerical values of $s$ and $t'$ into the
$9\times9$ mass matrix of Eq.~\eqref{eq:Mu}, we can diagonalize
$\mathcal M^u$ (and likewise $\mathcal M^d$) explicitly. As our experimental
knowledge is limited to the three generations of light quarks, it is
sufficient to focus on the upper left $3\times 3$ blocks of both the full mass
matrices as well as the associated mixing matrices. Comparing these with the
input parameters, we find that the masses and mixing angles related to the
third generation are systematically to small. We therefore adjust the input
parameters such that the effective $3\times3$ Yukawa matrices of the full
theory match the matrices $Y_u$ and $Y_d$ of the SM. We denote these adjusted
input parameters as ``adapted flavour parameters'' (labelled by a tilde), and
define ranges (given in Table~\ref{tab:adapted}) over which we perform the scan.

\begin{table}[tb]
  \centering
  \small
  \begin{tabular}{cccccccc}
   \toprule &&&&&&&\\[-1.5ex]
     $\tilde m_u$ & $0.5 - 2.9 \,\MeV$   && $\tilde m_d$ & $1.2 - 4.8 \,\MeV$
   && $\tilde \theta^L_{12}$ & $12.89^\circ - 13.19^\circ$ \\[1mm]
     $\tilde m_c$ & $0.53 - 0.71 \,\GeV$ && $\tilde m_s$ & $30 - 78 \,\MeV$
   && $\tilde \theta^L_{23}$ & $1.54^\circ - 2.56^\circ$ \\[1mm]
     $\tilde m_t$ & $162 - 288\,\GeV$    && $\tilde m_b$ & $2.78 - 4.44\,\GeV$
   && $\tilde \theta^L_{13}$ & $0.101^\circ - 0.28^\circ $  \\[1ex]  
\bottomrule
  \end{tabular}
  \caption{Ranges of the adapted flavour parameters over which the scan is
    performed. The mixing angles $\tilde \theta'^R_{ij}$ of
    $U'_\text{CKM}$ are varied in their full range $[0^\circ,90^\circ]$ and
    additionally in a reduced range  $[0^\circ,1.5^\circ]$. All phases are
    varied between $0$ and $2\pi$.}
  \label{tab:adapted}
\end{table}


\subsection{Details of the scan}

Having defined the adapted flavour parameters we describe in detail
our systematic scan over possible flavour effects. We do not claim this scan
to be exhaustive as the procedure may systematically exclude allowed regions
of the underlying parameter space. Nevertheless, it provides important
insights into the essential flavour effects of this model. The scan itself is
performed using the following steps.
\begin{enumerate}
\item Randomly generate a point in the space of adapted flavour parameters
   where the allowed ranges are given in Table~\ref{tab:adapted}. 
  Furthermore, choose also random values for $\lambda\in [1.5,3]$,
  $\tan\beta\in[1,15]$ and  $M\in[750,2500]\,\text{GeV}$.
\item Calculate $s$ and $t'$ by inverting Eq.~\eqref{eq:Yukrelations} using
  the procedure described in Appendix~\ref{sec:seesaw}. 
Here we randomly choose one of the eight solutions.
\item Insert the so-derived $s$ and $t'$ as well as $\lambda$, $\tan\beta$
   and $M$ into the full $9\times9$ mass matrices $\mathcal M^{u,d}$ and
   diagonalize these numerically.
\item Deduce the effective SM-like Yukawa parameters for the three light generations,
   i.e.\ $m_i^{u,d}$ and $V^{\text{eff}}_{\text{CKM}}$. 
\item Compare these with the experimentally allowed
   $3\sigma$~ranges.\footnote{We consider the masses at the electroweak scale,
  i.e.\ $m_i^{u,d}(M_Z)$~\cite{Xing:2007fb}, and the experimental constraints
  on the absolute 
  values of the individual CKM entries (not those obtained from additionally
  demanding unitarity of the CKM matrix).}  
  In case of agreement, save the point $(s, t', \lambda, \tan\beta, M)$ as a
  viable choice of input parameters. 
\end{enumerate}
Following these steps, we have generated about 3,000 viable points using the full
range of adapted parameters. However, scanning the full parameter space of the
mixing angles $\tilde \theta'^R_{ij}$ of $U'_\text{CKM}$ is rather
inefficient as only a small fraction of all points turns out to be physically acceptable.
For this reason, we have additionally performed a more extensive scan where we
choose random values of $\tilde \theta'^R_{ij}$ within the small interval
$[0^\circ,1.5^\circ]$. 
The resulting 30,000 viable points differ qualitatively from the full scan only in
$U_\text{CKM}$, i.e.\ the coupling of the right-chiral quarks to the $W^\pm$ bosons. 
In order to enhance the statistics, we have combined both sets of points in
our analysis (unless we are interested in the results for $U_\text{CKM}$ itself).

Before turning to the presentation and discussion of the physical results of our
scan, we comment on the qualitative structure of the flavon VEVs $\VEV S$ and
$\VEV {T'}$. In the original GRV model~\cite{Grinstein:2010ve}, the flavon
VEVs can be  derived from the SM Yukawa matrices via the approximate relation
$\VEV S\propto Y^{-1}$. This simple relation entails that the flavon VEVs
directly inherit the hierarchies of the Yukawa matrices in an inverse way. In
contrast to this, the situation in our PS flavour model is more
involved. Due to the linear combination of the VEVs $\VEV S$ and $\VEV{T'}$ in
Eq.~\eqref{eq:Yukrelations}, a simple direct relation of the hierarchies in
the SM Yukawas and the flavon VEVs does not exist. While many viable points of
the scan do indeed feature flavon VEVs with hierarchies similar to those of
the inverse Yukawas, there exists a significant number of points where the
hierarchies of VEVs is less (or more) pronounced. Similarly, the mixing angles
describing the mismatch of the matrices $s$ and~$t'$ can in some cases be
significantly larger than those of the CKM matrix.




\section{Quark flavour phenomenology}
\label{sec:GRVQuarkPheno}
\cleqn

In this section, we discuss the physical implications of our PS flavour model. 
Starting from the adapted flavour parameters together with the
ranges for $\lambda$, $\tan\beta$ and~$M$ as defined in
Section~\ref{sec:scan}, we observe that the values of $\tan\beta$ and~$M$
corresponding to the viable points are equally distributed throughout the full
allowed ranges. In contrast to this, the parameter $\lambda$ does not cover the
full allowed range but only $\lambda \in [1.5\,,\,2.5]$. In fact, it peaks around
$\lambda=2$, which is exactly the value we have used to determine the ranges
of the adapted flavour parameters. In principle, we could have expanded our
scan by additionally deriving adapted flavour parameters for other values of
$\lambda$, thereby enlarging the intervals in Table~\ref{tab:adapted}. 
However, we refrain from doing so as we are mainly interested in this part of the
parameter space.\footnote{In the one generation case it is possible to show that
 $\lambda \gtrsim \sqrt{2}$, and larger values are disfavoured in a perturbative theory.}

Regarding the $M$-dependence of all physical results, we note that this has
been factored out in the definition of the flavour breaking VEVs of
Eq.~\eqref{eq:VEVs}. Following this parameterization, the approximate Yukawa
relations of Eq.~\eqref{eq:Yukrelations} do not explicitly depend on~$M$.
Likewise, we do not expect any explicit $M$-dependence of the $3\times 3$ matrices
$a\ldots f$, cf.~Eq.~\eqref{eq:Mu3}. On the other hand, an implicit
$M$-dependence may in principle be induced via the intricate procedure of
determining $s$ and $t'$, in particular for the third generation of quarks. 
We have therefore checked for such a possibility in our scan. However, no
extra (implicit) dependence on $M$ is found in any of the quantities, at least
above a threshold of $M\sim 1\,\TeV$. This observation allows us to understand 
the dependence on the scale $M$ explicitly throughout the model, particularly
in all corrections.

As an application of this fact, we may limit our discussion of characteristic
flavour effects to a narrower mass band. In some of the following studies we
choose this to be $M\in[1,1.2]\,\TeV$, which reduces our set to approximately $3,500$
viable points. This restriction clarifies the dependence of physical
observables on the other input parameters of the scan as it reduces the spread due
to the variation of $M$. The results obtained for this reduced mass band may
afterwards be used to extrapolate the effects to larger values of $M$ using
the explicitly known $M$-dependence. Reversely, we can also use the explicit
$M$-dependence to rescale all generated points to a single mass
scale. We make use of such a rescaling in the discussion of $U_\text{CKM}$
in Section~\ref{sec:Uckm} in order to increase the statistics.


\subsection{Effective SM flavour parameters}

When defining the ranges of the adapted flavour parameters, we have aimed at
choices which entail a coverage of the complete range of experimentally
allowed Yukawa parameters. This turns out to be possible for most of the
parameters, however, not all of them are distributed equally over the allowed
range. 

Especially the absolute values of the CKM elements $|V_{tb}|$,
$|V_{ts}|$, $|V_{cs}|$ and $|V_{cd}|$ are more constrained in our setup. 
This can be traced back to fact that the direct bounds on these CKM elements
are relatively weak while the correction $\Delta V_\text{CKM}$ (related to
non-unitarity) is small. If we instead compare the covered range to the CKM
elements as deduced from the SM fit assuming unitarity~\cite{UTfit,CKMfitter},
we generate many points outside the $3\sigma$~region. In particular, we find
that our $|V_{tb}|$ is typically smaller. The experimental bounds we impose on the
individual CKM elements, the ranges deduced from the SM fit and the ranges
covered in our scan are all given in Table~\ref{tab:flavourcoverage} for
comparison.
\begin{table}[t]
  \centering
  \begin{tabular}{lcccccc}
\toprule &&&&&&\\[-1.5ex]
    &\phantom{X}& $|V_{ud}|$ && $|V_{us}|$ && $|V_{ub}|$   \\[.5ex]
    experiment  && $0.9736 - 0.9749$ &&$ 0.2229 - 0.2277$ && $0.0027 - 0.0056$ \\
    unitarity   &&$ 0.9739 - 0.9747 $&& $0.2235 - 0.2272 $&& $0.0031 - 0.0040$ \\
    covered     &&$ 0.9737 - 0.9748 $&& $0.2231 - 0.2277 $&&$ 0.0027 - 0.0056$  \\[3ex]
    && $|V_{cd}|$ && $|V_{cs}|$ && $|V_{cb}|$ \\[.5ex]
    experiment  && $0.2010 - 0.2490 $&& $0.9380 - 1.0340 $&&$ 0.0372 - 0.0450$ \\
    unitarity   &&$ 0.2234 - 0.2271 $&& $0.9730 - 0.9739 $&&$ 0.0378 - 0.0450$ \\
    covered     &&$ 0.2229 - 0.2276 $&&$ 0.9727 - 0.9741 $&&$ 0.0372 - 0.0450$  \\[3ex]
    && $|V_{td}|$ && $|V_{ts}|$ && $|V_{tb}|$ \\[.5ex]
    experiment  &&$ 0.0066 - 0.0102 $&&$ 0.0319 - 0.0481$ && $0.9250 - 1.1170$ \\
    unitarity   && $0.0079 - 0.0099 $&& $0.0372 - 0.0438 $&& $0.9990 - 0.9993$ \\
    covered     && $0.0066 - 0.0102 $&&$ 0.0355 - 0.0447 $&&$ 0.9711 - 0.9992$ \\[1ex]
\hline
  \end{tabular}
  \caption{Coverage of the allowed range of
    $\big|\left(V^{\text{eff}}_{\text{CKM}}\right)_{ij}\big|$. The first line
    corresponds to the direct experimental limits we impose for the
    scan~\cite{Agashe:2014kda}. The second line shows the allowed range
    deduced from the SM fit assuming unitarity of the CKM
    matrix~\cite{Agashe:2014kda}. The last line corresponds to the range covered in our scan.}
  \label{tab:flavourcoverage}
\end{table}
Concerning the masses $m^{u,d}_i$ we find that they all cover the full allowed range~\cite{Xing:2007fb}.
However, there is a tendency to lower values especially for the charm mass.
For the light quark masses, we also see a deficit in the largest values of the
allowed ranges, although not as pronounced as for the charm mass.
We interpret these effects as relics of our scan, as the setup of the model
generally lowers the masses. The allowed ranges for the masses can
be found in Table~\ref{tab:allowedrange}.
\begin{table}[t]
  \centering
  \small
  \begin{tabular}{lcclcclc}
   \toprule &&&&&&&\\[-1.5ex]
     $m_u$ & $0.5 - 2.9 \,\MeV$   && $m_d$ & $1.2 - 4.8 \,\MeV$   
\\
     $m_c$ & $0.53 - 0.71 \,\GeV$ && $m_s$ & $30 - 78 \,\MeV$     
 \\
     $m_t$ & $162 - 180\,\GeV$    && $m_b$ & $2.78 - 2.96 \,\GeV$ 
\\[1ex]
\bottomrule
  \end{tabular}
  \caption{Ranges of the quark masses $m^{u,d}_{i}$ covered in the scan.}
  \label{tab:allowedrange}
\end{table}

Having checked  that the model is capable of generating the flavour structure
of the SM, we now turn our attention to quark flavour effects beyond the
Standard Model.


\subsection{Masses of the new heavy quarks}

The masses of the heavy up-type quark partners can be read off from the
approximately diagonalized mass matrix of Eq.~\eqref{eq:Mu4}. A similar expression
can be deduced for the down-type quark partners by replacing the index $u$ by $d$.
As we have explicitly factored out the mass scale $M$  it is
reasonable to consider only the mass ratio 
\begin{align}
{\mu}_F = \frac{m_F}{M}\,.
\end{align}
With the matrix $\hat{f}$ being independent of isospin, we can directly infer
that one set of heavy up-type partners equals one set of down-type partners in
mass, denoted by $U''_i$ and $D''_i$ in the following, where $U_i$ ($D_i$) are the
three generations of up-type (down-type) quarks. This observation is
numerically verified in our scan, where the corresponding masses are equal at
the sub-percent level. Introducing a VEV $\VEV{T}\neq0$ for the
$SU(2)$ triplet flavon~$T$ would break this mass degeneracy by an amount
proportional to $\VEV{T}$. Being proportional to $\hat e_{u,d}$, the masses of
the other set of heavy quark partners, denoted by $U'_i$ and $D'_i$, will
generally depend on isospin.

\begin{figure}[tb]
\begin{minipage}[b]{.48\textwidth}
 \centering
 \subfigure[First solution.]{\includegraphics[width=.95\textwidth]{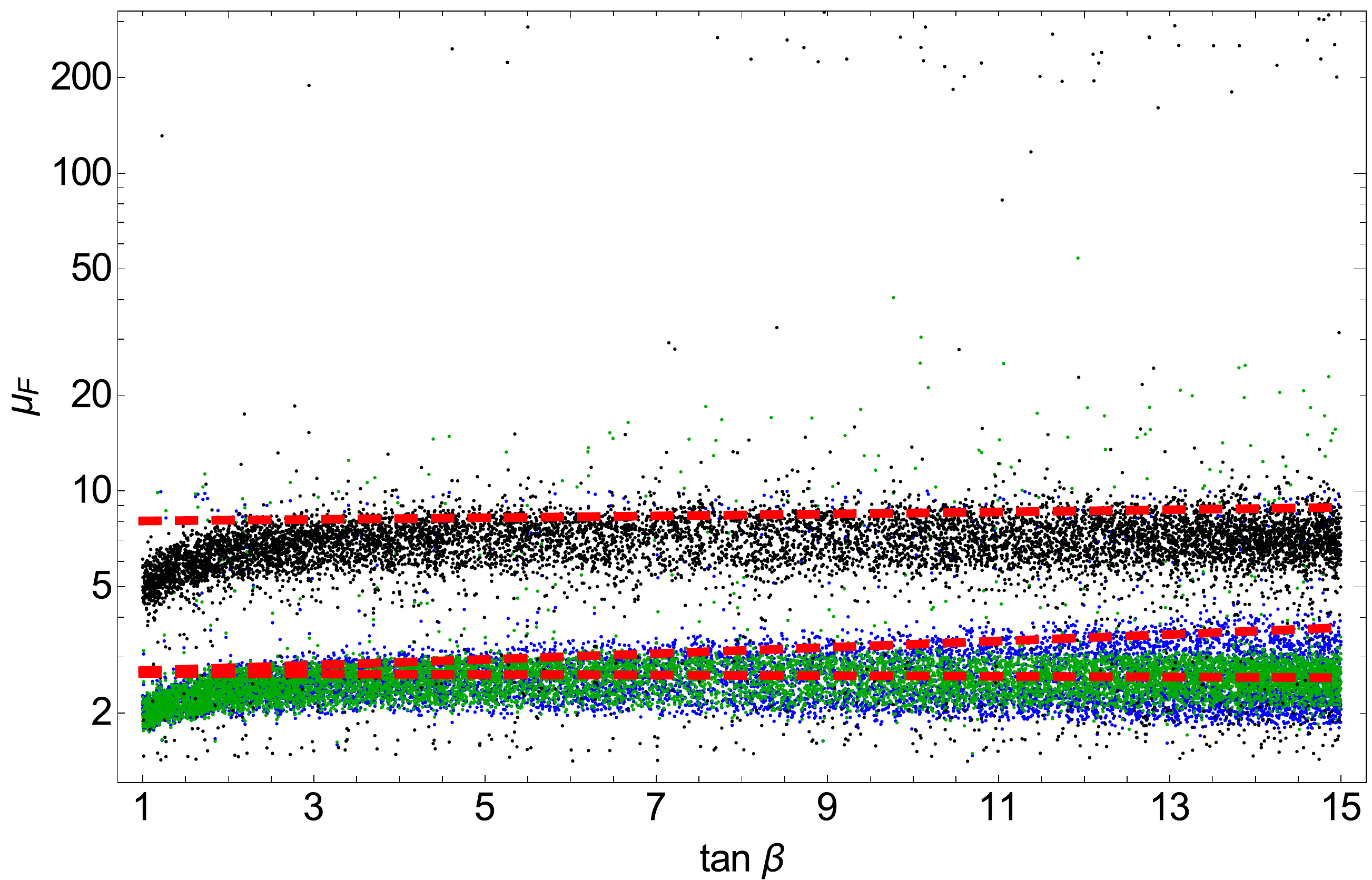}\label{fig:GRVmasses1}}
\end{minipage}\hfill
\begin{minipage}[b]{.48\textwidth}
 \centering
 \subfigure[Second solution.]{\includegraphics[width=.95\textwidth]{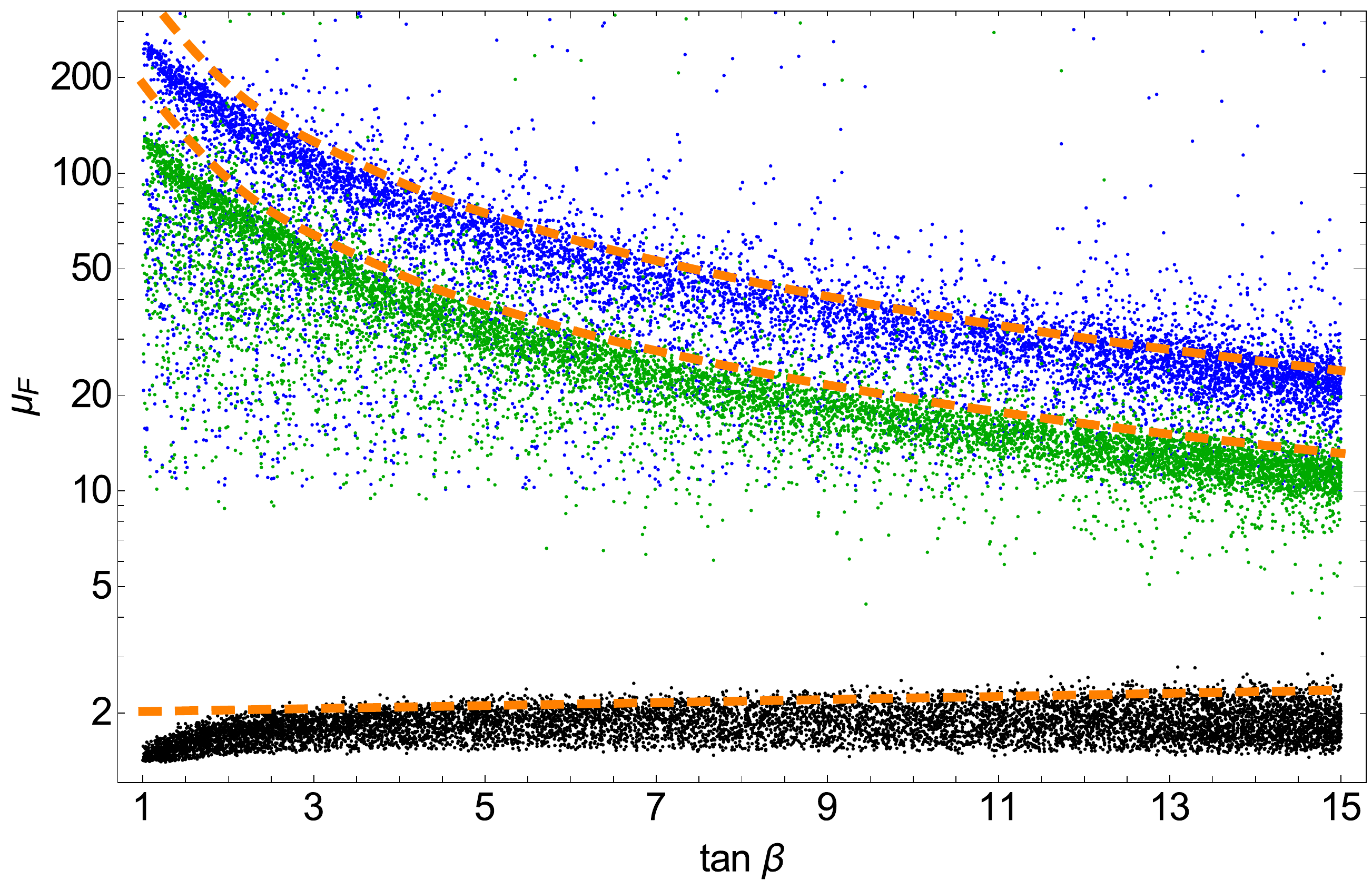}\label{fig:GRVmasses2}}
\end{minipage}
\caption{Mass ratios ${\mu}_F$ of the lightest generation of heavy quark
  partners with respect to $\tan\beta$. The plots show the masses of 
the first top partner in black, the first bottom partner in blue,
and the second top and bottom partners in green. 
The two solutions of Eq.~\eqref{eq:GRVInversion} are given for each mass ratio
as dashed lines in red (a) and orange (b), respectively.}
\label{fig:addmasses}
\end{figure}

Focusing on the lightest generation of heavy quark partners, i.e.\ $t'$, $b'$, $t''$
and $b''$, the multiplicity of solutions to Eq.~\eqref{eq:Yukrelations} is
dominated by the multiplicity of the one generation case, see
Eq.~\eqref{eq:GRVInversion}. The scan therefore generates two distinct bands for
each of the four mass ratios~$\mu_F$. These in turn depend on $\tan\beta$,
which can be traced back to the appearance of $y_u$ and $y_d$ in
Eq.~\eqref{eq:GRVInversion}. The resulting $\tan\beta$-dependence of the heavy
quark masses calculated from the one generation case fits the results of the
scan reasonably well as illustrated in Figure~\ref{fig:addmasses}. 
Here, we have plotted the mass ratios ${\mu}_F$ of the four lightest
quark partners ($t'$, $b'$, $t''$ and $b''$) against $\tan\beta$.
For clarity we have separated the two possible solutions. The mass ratios of
$t''$ and $b''$ are shown in green (as both are equal), the one of $t'$ in
black and that of $b'$ in blue.
Additionally, we have plotted the $\tan\beta$-dependence of the mass ratios
deduced from Eq.~\eqref{eq:GRVInversion} where we have used the central values of 
the SM Yukawa couplings and $\lambda=2$ as fixed input.
Here, we introduce a colour coding for this and the following plots: masses
belonging theoretically to the first (second) solution of
Eq.~\eqref{eq:GRVInversion} are shown in red (orange). 

We can conclude from the plots of Figure~\ref{fig:addmasses} that the mass
ratios $\mu_F$ of the heavy quark partners may be of order one, and thus their
masses can be in the $\TeV$-regime. However, depending on the solution, we may
expect either a set of three quark partners ($b'$, $t''$ and~$b''$) within a
small mass range or solely the top partner~$t'$ to appear in
experiments. This shows that the two solutions differ qualitatively in their
phenomenological predictions. We will encounter further differences in the
following discussions.


\subsection[Non-unitarity~of~the~effective~CKM~matrix]{Non-unitarity of $\boldsymbol{V}^{\text{eff}}_{\text{CKM}}$}

Due to the $\epsilon^2_{u,d}$-dependent contribution 
$\Delta V_\text{CKM}$ to the effective CKM matrix, see 
Eq.~\eqref{eq:GRVflavoureffects}, the coupling of the $W^\pm$ bosons to the
three generations of light quarks is no longer described by a unitary matrix.
This non-unitarity can be quantified by studying the unitarity triangles
constructed from ${V}^{\text{eff}}_{\text{CKM}}$. For the sake of clarity, we
limit our discussion to the ``standard unitarity triangle''. However, we have
checked that the other unitarity triangles lead to analogous results with
similar or smaller effects.

Generically we do not expect large deviations from the SM as we have limited
our scan to points which are compatible with the absolute values of the CKM
matrix elements, taking into account the uncertainty of the measurements.
This is a simplifying assumption as the determination of the CKM elements from
experimental measurement is also affected by the couplings of the right-chiral
quarks. To give a complete picture, we would have to redo the extraction
of the CKM elements in the presence of additional couplings of right-chiral
quarks ($U_\text{CKM}$) and without the assumption of unitarity. Clearly,
such an explicit fit is beyond the scope of our current work.

\begin{figure}[tb]
\begin{minipage}[b]{.48\textwidth}
 \centering
 \subfigure[The origin of the standard unitarity triangle. The blue-shaded
   regions correspond to the $1\sigma$ and $3\sigma$ ~uncertainty of the
   measured SM Wolfenstein parameters~$\overline{\rho}$ and
   $\overline\eta$~\cite{Agashe:2014kda}.]
{\includegraphics[width=.95\textwidth]{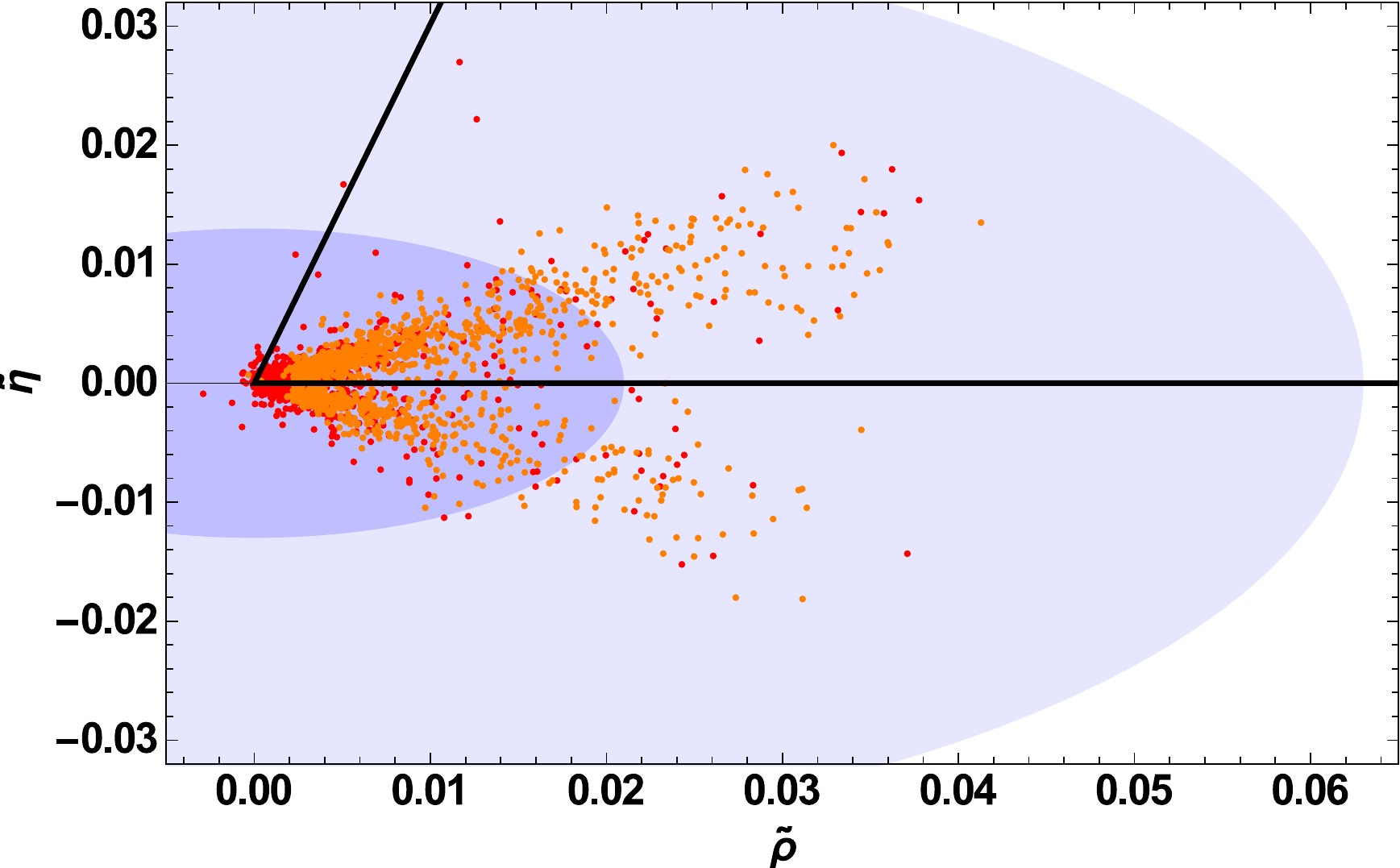}\label{fig:utriangle}}
\end{minipage}\hfill
\begin{minipage}[b]{.48\textwidth}
 \centering
 \subfigure[The absolute value of $\widetilde{\rho}+i\,\widetilde \eta$ plotted against
   $\tan\beta$ for points with $M\in \text{[1,1.2]}\,\TeV$. The dashed blue line
   indicates the corresponding $1\sigma$~error of
   $|\overline{\rho}+i\,\overline{\eta}|$ as obtained in the SM~\cite{Agashe:2014kda}.]
{\includegraphics[width=.95\textwidth]{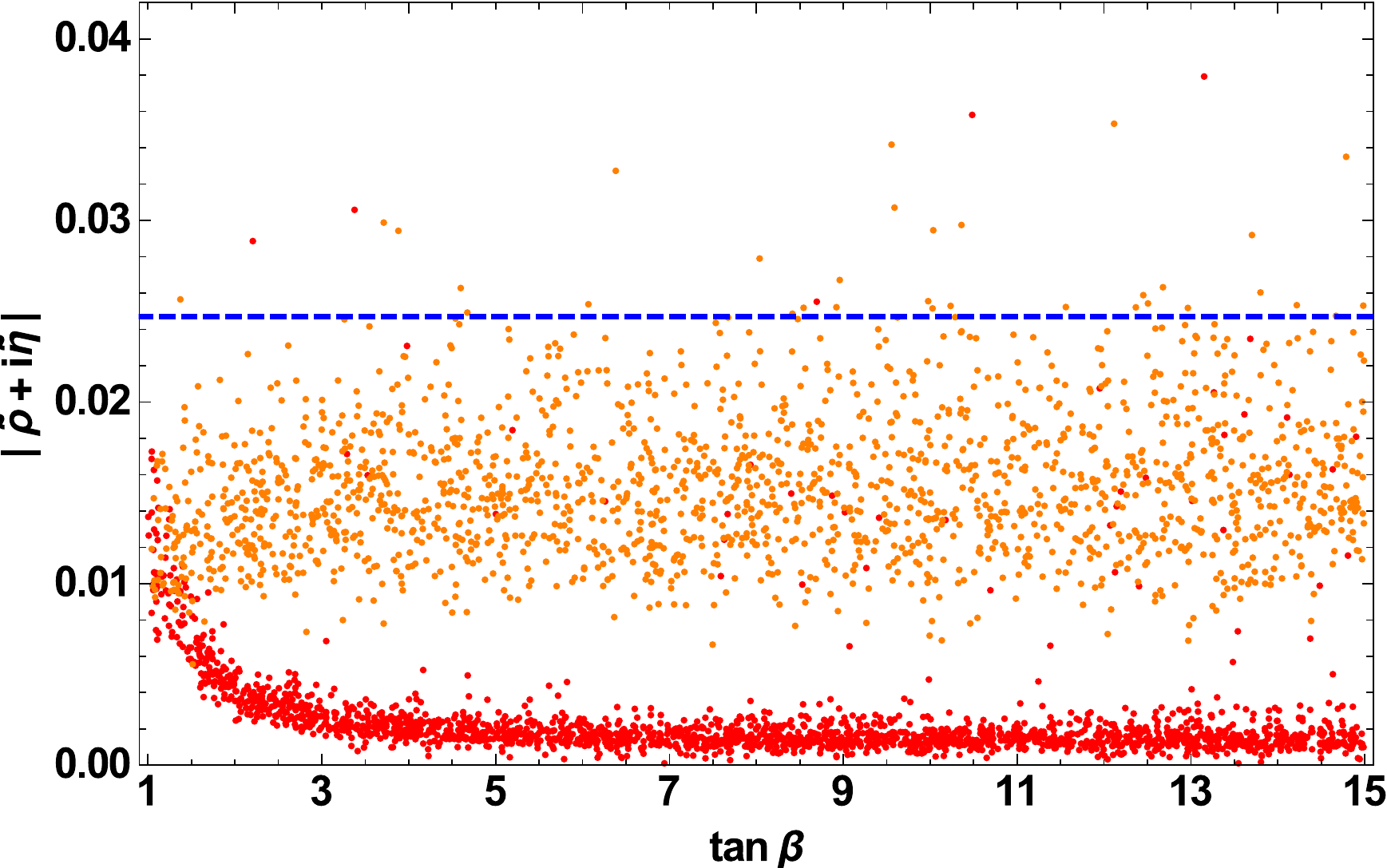}\label{fig:VCKMununitarity}}
\end{minipage}
\caption{The non-unitarity of ${V}^{\text{eff}}_{\text{CKM}}$. For the
  first (second) solution of Eq.~\eqref{eq:GRVInversion}, we 
  show~$\widetilde{\rho}+i\,\widetilde \eta$, i.e.\ the measure of the
  deviation from unitarity, in red (orange).} 
\label{fig:unitaritytriangle}
\end{figure}

In order to quantify the non-unitarity of ${V}^{\text{eff}}_{\text{CKM}}$ via
the standard unitarity triangle, we define the complex quantity  \vspace{-1.5mm}
\begin{align}
  \widetilde{\rho}+i\,\widetilde \eta~ \equiv ~\frac{V_{ub}V_{ud}^{*}+V_{cb}V_{cd}^{*} +V_{tb}V_{td}^{*}}{V_{cb}V_{cd}^*} \ ,
\end{align}\\[-4.5mm]
whose deviation from zero is a measure of the deviation from unitarity.
In Figure~\ref{fig:unitaritytriangle}, we show $\widetilde \rho+i\,\widetilde \eta$ for the
complete set of viable points at the origin of the unitarity triangle. 
In addition, we present the $\tan\beta$-dependence of its absolute value
$|\widetilde{\rho}+i\,\widetilde \eta|$, where we limit ourselves to viable points which lie in
the mass band $M\in[1,1.2]\,\TeV$ for the sake of clarity.
The points associated with the two solutions are again distinguished by the
colour code introduced above. We compare these results of our scan with the
$1\sigma$ and $3\sigma$~uncertainties of the measured SM Wolfenstein
parameters $\overline{\rho}$ and $\overline{\eta}$~\cite{Agashe:2014kda}.
As expected, the deviation from unitarity predicted in our scan is
within the SM uncertainties. Moreover, we again find that the size of this
effect depends crucially on the type of solution. This is particularly
apparent in the $\tan\beta$-dependence of $|\widetilde{\rho}+i\,\widetilde\eta|$,
cf. Figure~\ref{fig:VCKMununitarity}. Here, the second (orange) solution is
basically independent of $\tan\beta$ whereas the first (red) is of similar
size for $\tan\beta=1$, but becomes negligible for $\tan\beta\gtrsim3$.
Thus, the effect can be practically absent even for small values of the
flavour breaking mass scale $M$. Considering the values of 
$\widetilde \rho+i\,\widetilde\eta$ in the complex plane near the origin of
the unitarity triangle, we observe that the real part $\widetilde{\rho}$ is
generically positive, 
cf. Figure~\ref{fig:utriangle}. This indicates that the effect is dominated by
the decrease of $|V_{tb}|$ which in turn reduces the length of the upper right
line of the unitarity triangle. 


\subsection[Coupling~of~right-chiral~quarks~to~W~via~UCKM]{Coupling of right-chiral quarks to $\boldsymbol{W^\pm}$ 
via $\boldsymbol{U}_{\text{CKM}}$}
\label{sec:Uckm}  

\begin{figure}[tb]
\begin{minipage}[b]{.48\textwidth}
 \centering
 \subfigure[Impact of $U_{ub}$ on the determination of $|V_{ub}|$ as given
   in~\cite{Crivellin:2014kga}. The coloured 
   bands show different measurements: inclusive decays (blue),
   $B\rightarrow\pi l\nu$ (red), $B\rightarrow\rho l\nu$ (yellow),
   $B\rightarrow\tau\nu$ (green). The points obtained in our scan are
   shown as black dots.] 
{\includegraphics[width=.95\textwidth]{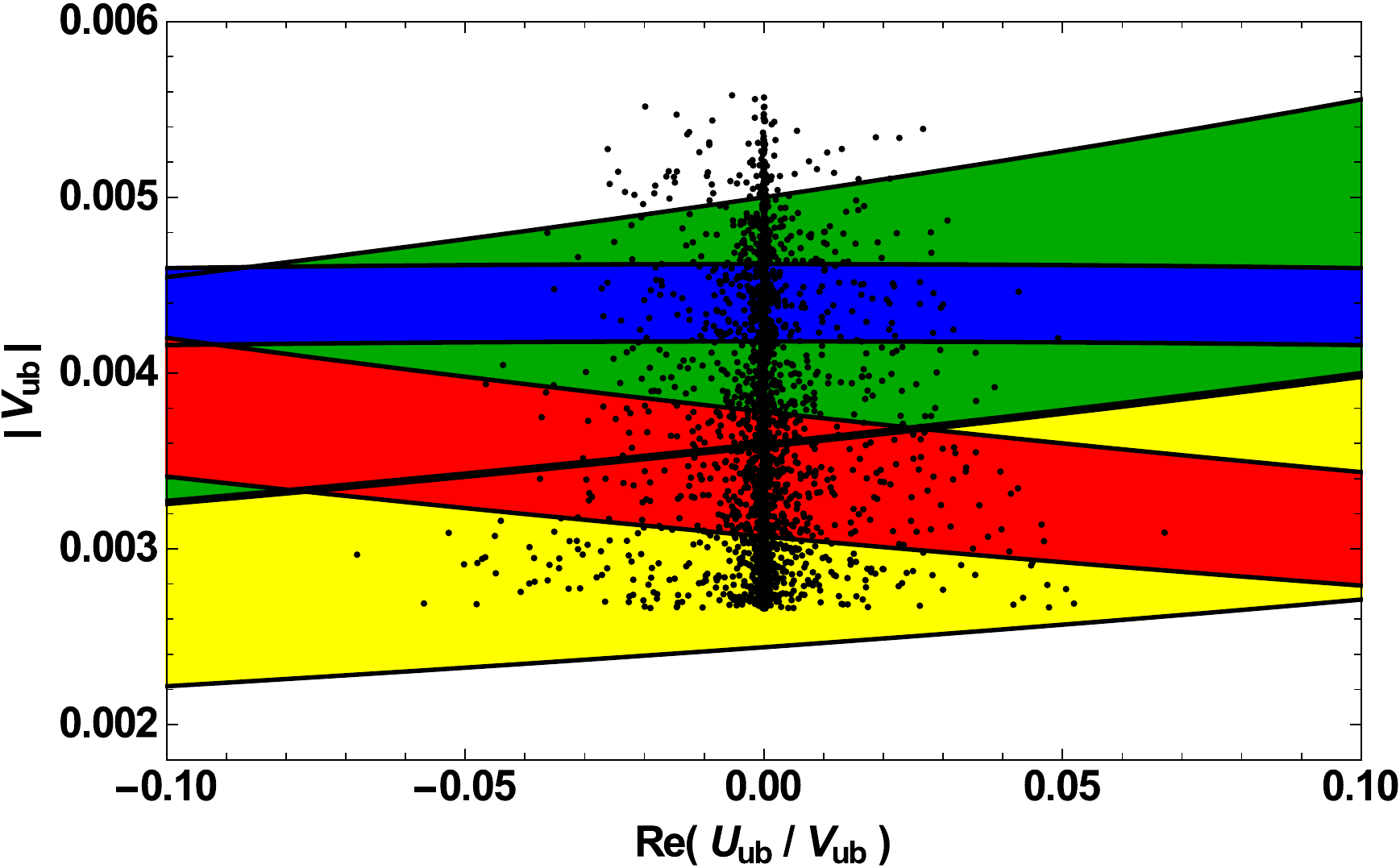}\label{fig:GRVVpub}}
\end{minipage}\hfill
\begin{minipage}[b]{.48\textwidth}
 \centering
 \subfigure[Impact of $U_{cb}$ on the determination of $|V_{cb}|$ as given
   in~\cite{Crivellin:2014kga}. The coloured
   bands show different measurements: inclusive decays (blue),
   $B\rightarrow D^* l\nu$ (red), $B\rightarrow D l\nu$ (yellow). The points
   obtained in our scan are shown as black dots.]
{\includegraphics[width=.95\textwidth]{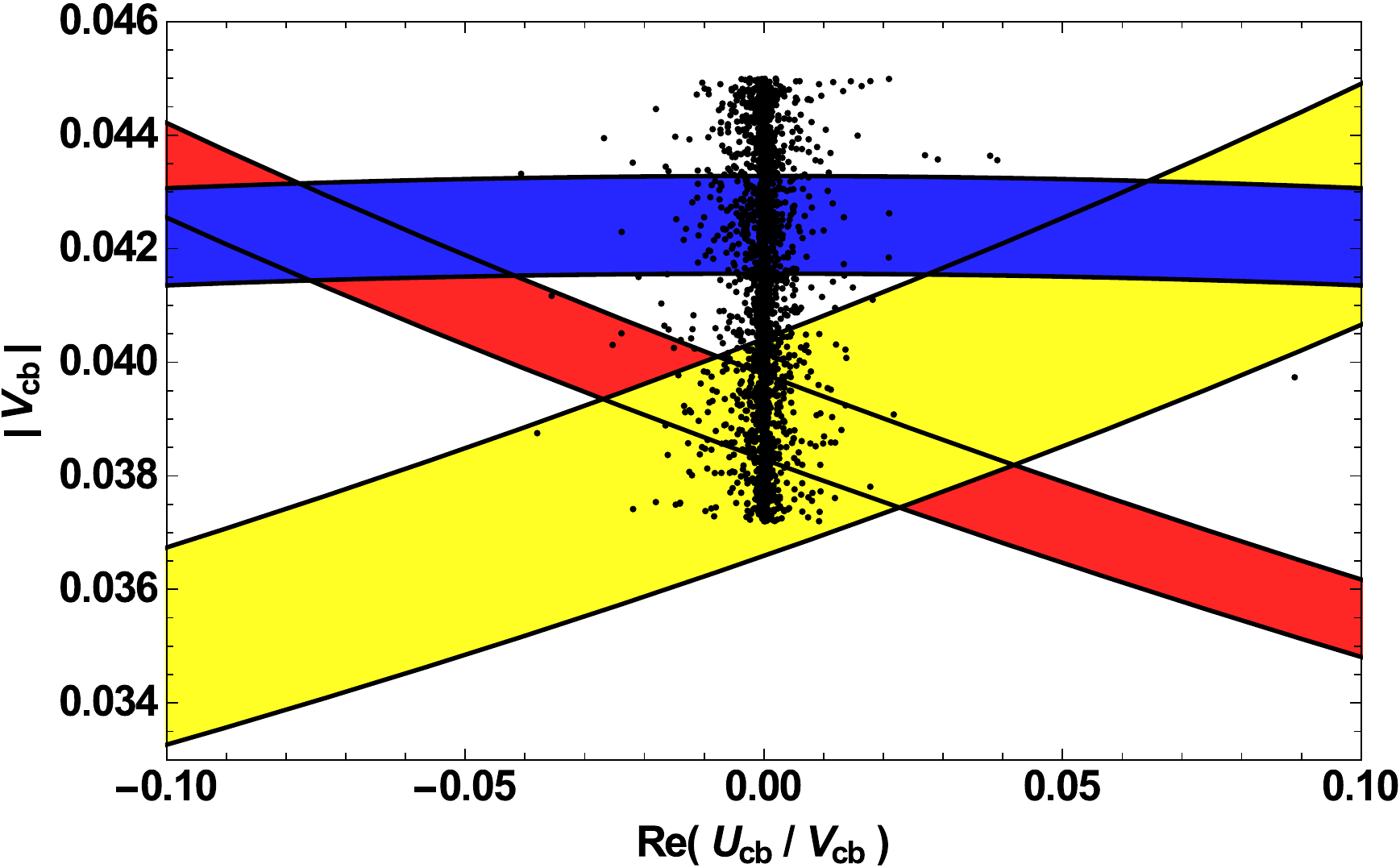}\label{fig:GRVVpcb}}
\end{minipage}
\caption{The impact of $U_\text{CKM}$ on the determination of
  $|V_\text{CKM}|$. Using the explicit $M$-dependence, the points of 
  our scan have been scaled to the reference scale $M=1\,\TeV$.} 
\label{fig:GRVVCKMp}
\end{figure}

As described in Section~\ref{sec:gaugekin}, the light generations of right-chiral
quarks couple to the SM $W^\pm$ gauge boson due to the mixing of $q_R$ with
$\Xi_R$ in Eq.~\eqref{eq:fermionlagrangian}. The corresponding coupling
strength is parameterized by $U_\text{CKM}$ as stated explicitly in
Eq.~\eqref{eq:GRVCKMp}. Formally it is proportional to~$\epsilon^2_{u,d}$ which we find
confirmed in our scan. Furthermore, we see no direct correlation between
$U_\text{CKM}$ and $V_\text{CKM}$.
The existence of such a coupling modifies the determination of
the CKM elements from experimental data. In particular, it might be
significant in view of the puzzling tension between the 
determination of $|V_{ub}|$ and $|V_{cb}|$ from exclusive and inclusive $B$-meson
decays within the SM~\cite{Agashe:2014kda,Bevan:2014iga}.
The additional couplings $U_{ub}$ and $U_{cb}$ of right-chiral quarks to
$W^\pm$ can in principle reduce (though not completely resolve)
this tension as the inclusive decay is proportional to
$|V_{xb}|^2 + |U_{xb}|^2$, while the exclusive decay is proportional
to\footnote{Here the sign is determined by the currents (axial or vector)
mediating the decay.} $|V_{xb}\pm U_{xb}|^2$, see e.g.~the discussions
in~\cite{Buras:2010pz,Crivellin:2014kga,Feldmann:2015xsa}. 
Yet, from the point of view of our model, our scan may reveal which experimental 
measurements are ``theoretically preferred''.

In our scan, we have searched for such effects, but do not find a conclusive
answer. Figure~\ref{fig:GRVVCKMp} shows the $1\sigma$~regions of
$|V_{ub}|$ and $|V_{cb}|$ as a function of $U_{ub}$ and
$U_{cb}$~\cite{Crivellin:2014kga}. We have additionally plotted the $3,000$ points of
our scan over the full parameter range, scaled to the reference mass scale $M = 1\,\TeV$.
As a result we find that the relative contributions of $U_\text{CKM}$ are at
the percent level at most. Furthermore, the scan shows no preference for any
of the measurements, and no apparent correlation between $U_{ub}$ and $U_{cb}$
is seen.


\subsection[Anomalous~Z~coupling]{Anomalous $\boldsymbol{Z}$ coupling} 

\begin{figure}[tb]
\begin{minipage}[b]{.48\textwidth}
 \centering
 \subfigure[Correlation between the coupling of the $Z$ boson to the left- and
   right-chiral bottom quark. The dashed grey line indicates the diagonal.]
 {\includegraphics[width=.95\textwidth]{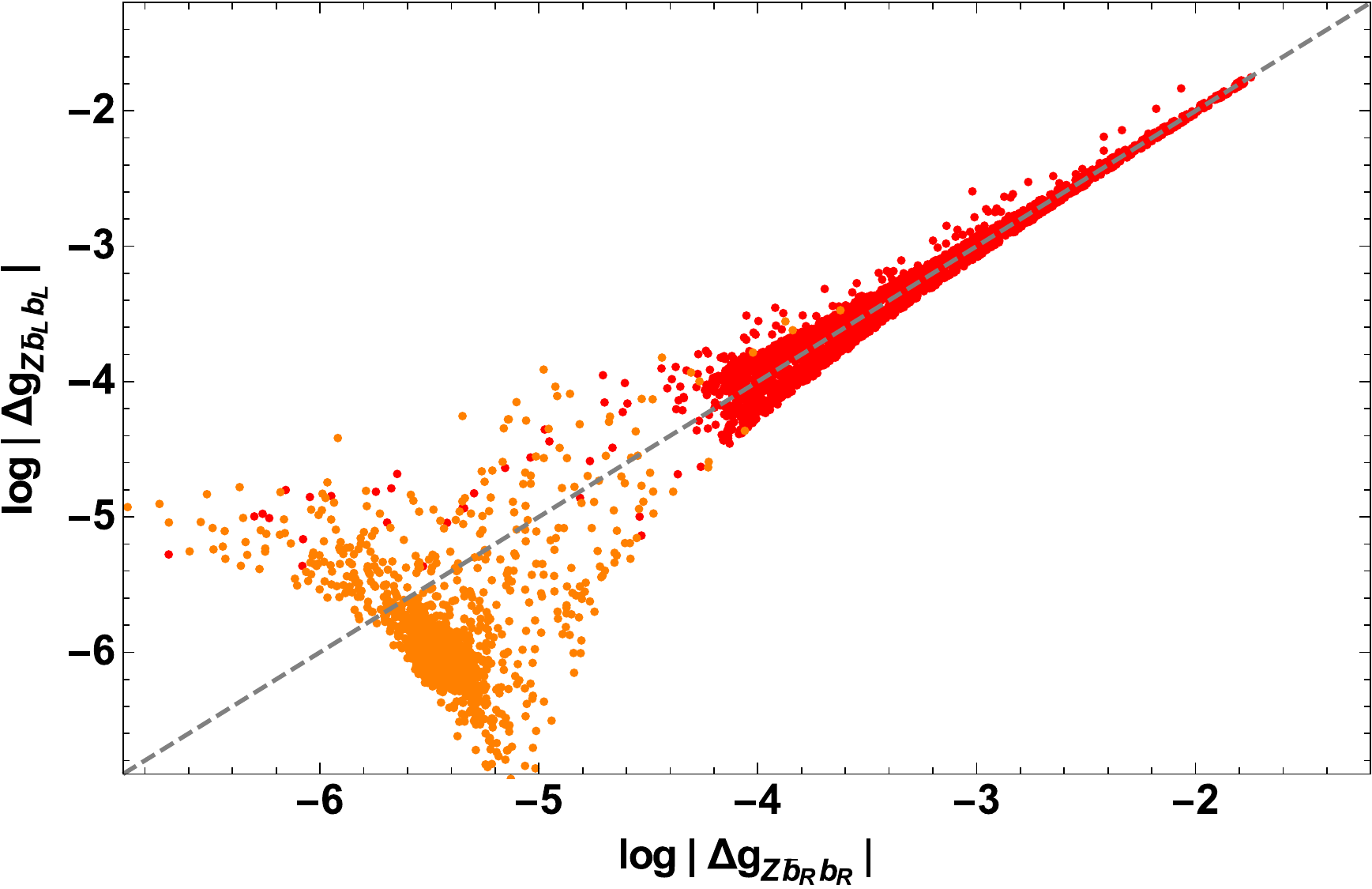}
 \label{fig:GRVZcorrelation}}
\end{minipage}\hfill
\begin{minipage}[b]{.48\textwidth}
 \centering
 \subfigure[The dependence of $\left|\Delta
   g_{Z\overline{b}^{}_Lb_L^{}}\right|$ on $\tan\beta$. The dashed blue line
   indicates the $3\sigma$  limit of Eq.~\eqref{eq:bound} (points below are allowed).] 
 {\includegraphics[width=.95\textwidth]{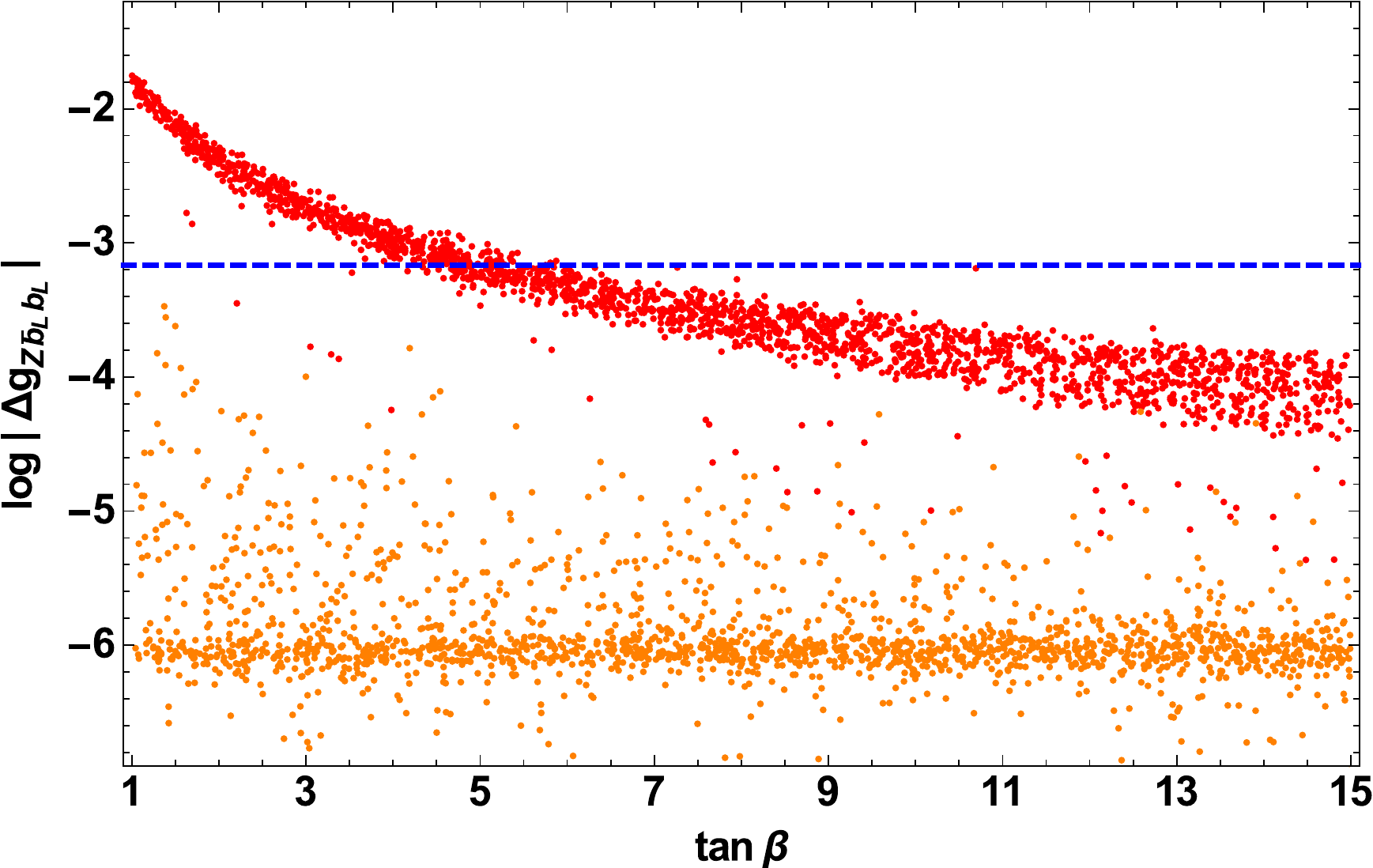}
\label{fig:GRVexcltanb}}
\end{minipage}
\caption{Correction of the $Z$ boson coupling to the $b$ quark. In red
  (orange) we show the points corresponding to the first (second) solution of
  Eq.~\eqref{eq:GRVInversion} for the mass band $M\in[1,1.2]\,\TeV$.}
\label{fig:excludedpoints}
\end{figure}

As can be seen explicitly in Eq.~\eqref{eq:L_kingauge}, the coupling of the
$Z$ boson to the left- and right-chiral quarks is modified as well. Similarly
to the effects discussed above, the largest deviation arises in the top quark
coupling, where it is in the range of a few percent for moderate values of~$M$. 
Experimentally, the $Z$-top-top coupling cannot be determined directly from the
decay of an on-shell $Z$ boson as it is kinematically forbidden. Thus top
physics processes such as $t\overline t Z$ production at the
LHC~\cite{Rontsch:2014cca}, or the production of a pair of top quarks mediated
by the neutral gauge bosons $Z^\ast$ and $\gamma^\ast$ at the
ILC~\cite{Rontsch:2015una} provide alternative and promising ways of
measuring the coupling of the $Z$ to the top quark.
However, as the resulting bounds are expected to be relatively weak, 
we focus on the coupling of the $Z$ to the bottom quark.
This decay has been measured precisely at LEP2~\cite{ALEPH:2005ab} 
with $\Gamma(Z\rightarrow \overline{b}b) \approx 375.87 \pm 0.17
\,\MeV$~\cite{Agashe:2014kda}.  
It involves the couplings of both the left- as well as the right-chiral
$b$ quarks. Considering the correlation of their corrections
within our model, plotted in Figure~\ref{fig:GRVZcorrelation}, we observe that
$\left| \Delta g_{Z\overline{b}_Lb_L}\right|$ and 
$\left| \Delta g_{Z\overline{b}_Rb_R}\right|$ 
are roughly of equal size, unless the values are small and
thus negligible for phenomenological purposes. We can therefore make the
simplifying assumption that both corrections are approximately identical.
Assuming furthermore that the correction to the $Z\overline b b$ coupling is
within the experimental uncertainty, we can set the following $3\sigma$~limit
on the couplings, 
\begin{align}\label{eq:bound}
 \left| \Delta g_{Z\overline{b}_Lb_L}\right| \sim \left|\Delta
 g_{Z\overline{b}_Rb_R}\right| \lesssim 6.8\times 10^{-4}   \ .
\end{align}
This bound reduces the number of viable points of the scan by
roughly 10 percent. The so-excluded points all belong to the first solution of
Eq.~\eqref{eq:GRVInversion}. Assuming $M\in[1,1.2]\,\TeV$,
Fig.~\ref{fig:GRVexcltanb} shows the $\tan\beta$-dependence of $\left| \Delta
g_{Z\overline{b}_Lb_L}\right|$ for both solutions. While the second (orange) solution
does not vary with $\tan\beta$, the first (red) is characterized by decreasing
values of $\left| \Delta g_{Z\overline{b}_Lb_L}\right|$ for increasing
$\tan\beta$. Thus, the bound of Eq.~\eqref{eq:bound} is satisfied more often
for larger values of $\tan\beta$,
and all model points corresponding to the first solution with $\tan\beta\lesssim4$ can be excluded 
(yet smaller values are allowed for larger $M$).


\subsection{Anomalous Higgs coupling} 
\begin{figure}[tb]
\begin{minipage}[b]{.48\textwidth}
 \centering 
 \subfigure[The normalized coupling $\gamma_{h\overline{t}t}$ as
   a function of~$M$. The SM expectation is shown by the dashed blue line. The
   error of the experimental direct measurement exceeds the plotted range.]
 {\includegraphics[width=.95\textwidth]{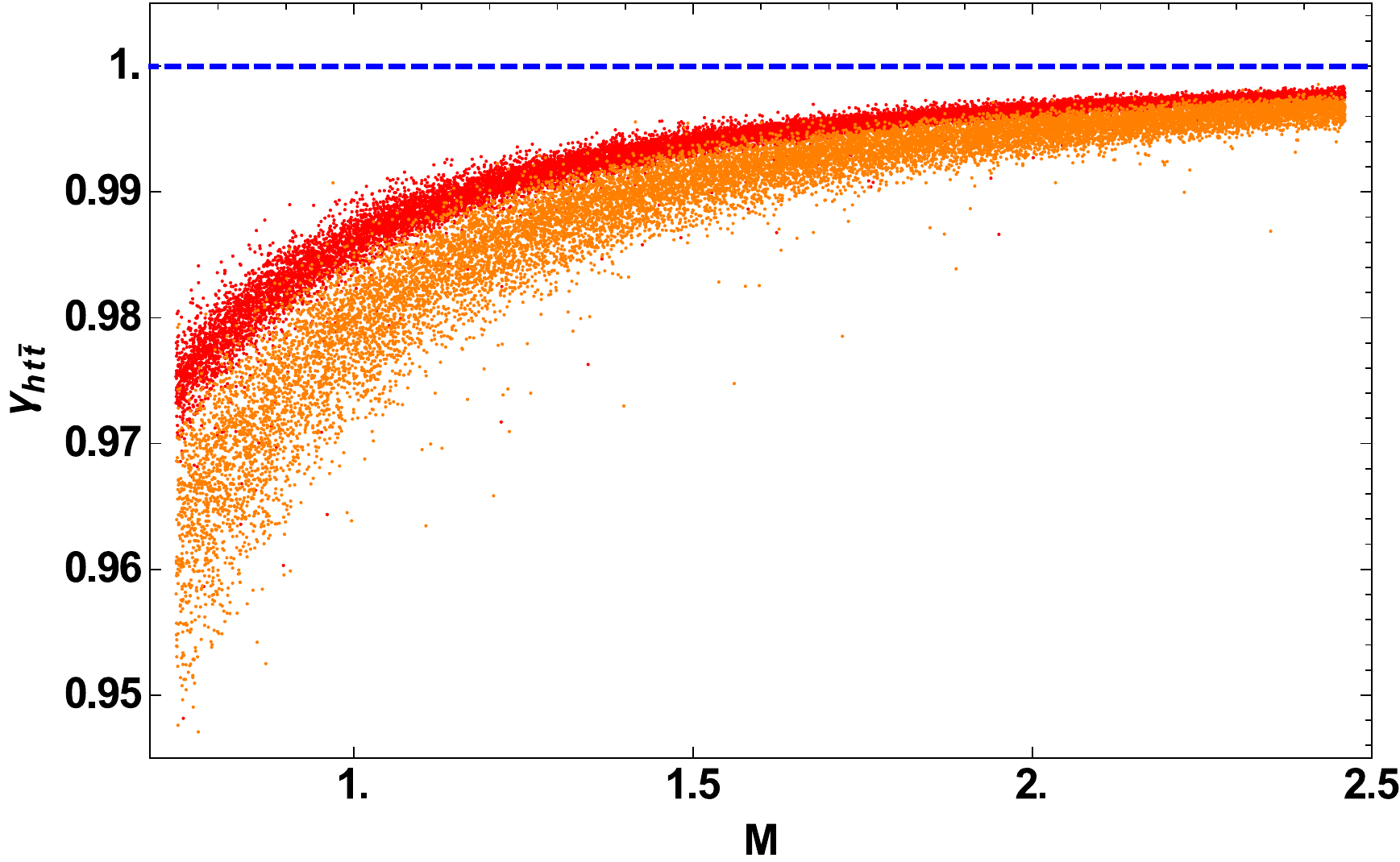}
 \label{fig:httvM}}
\end{minipage}\hfill
\begin{minipage}[b]{.48\textwidth}
 \centering
 \subfigure[Correlation of the normalized coupling $\gamma_{h\overline{t}t}$
   and $|V_{tb}|$. The point corresponding to the SM value is shown in
   blue. The experimental errors on both quantities exceed the plotted ranges.] 
 {\includegraphics[width=.95\textwidth]{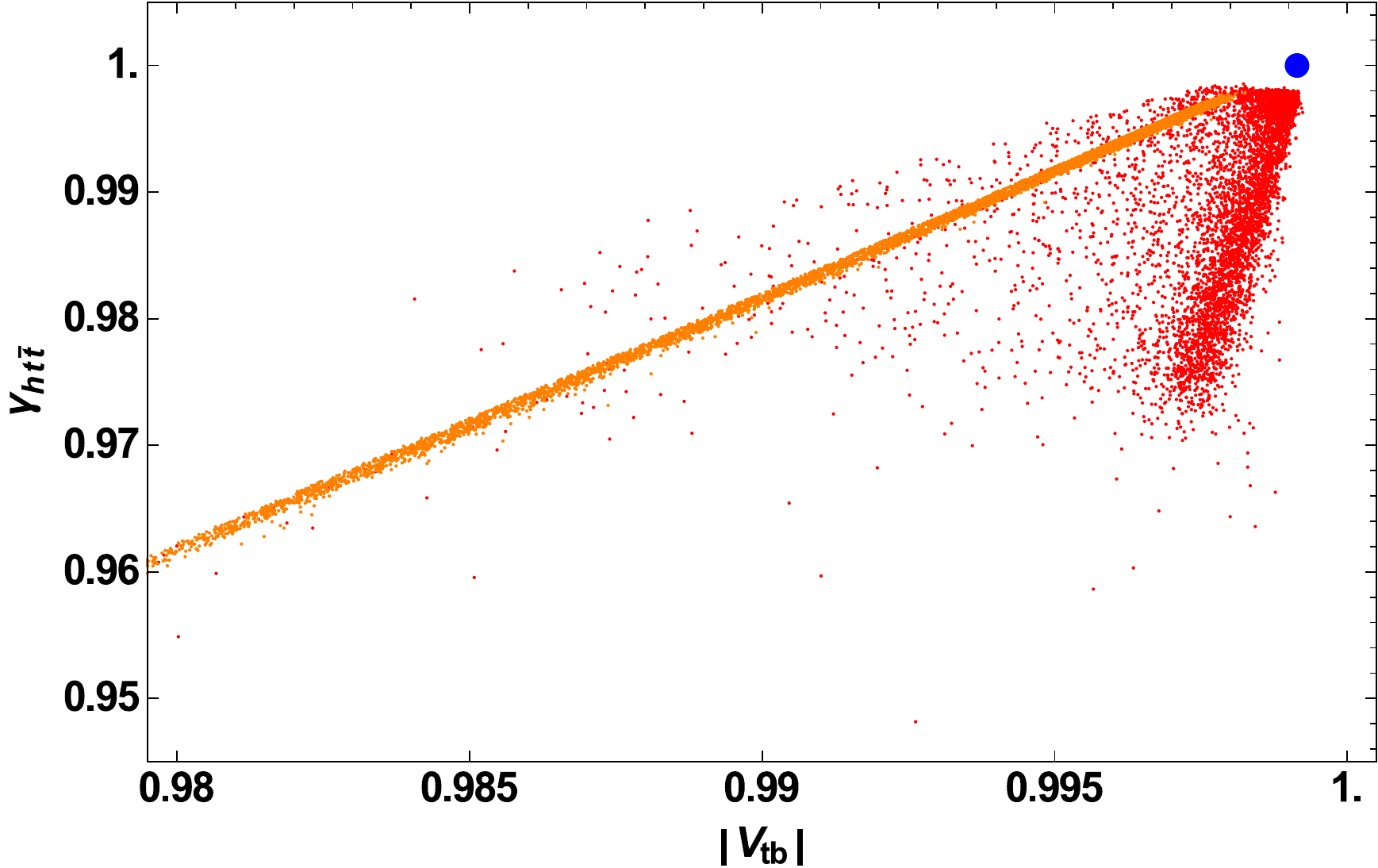}
\label{fig:Vtbvhtt}}
\end{minipage}
\caption{Correction of the normalized Higgs-top-top coupling
  $\gamma_{h\overline{t}t}$. In red (orange) we show the points corresponding
  to the first (second)  solution of Eq.~\eqref{eq:GRVInversion}.} 
\label{fig:fig:anomaloushiggs}
\end{figure}

Following the discussion of Section~\ref{sec:GRVHiggscoupling}, the Higgs
coupling to the top quark $g_{h\overline{t}t}$ can receive sizeable
corrections in our setup. However, such an effect is negligible already for
the $b$~quark as well as for all lighter quarks. Being of order $\epsilon^2_{u}$,
see Eq.~\eqref{eq:GRVhnlo}, the correction to $g_{h\overline{t}t}$ reduces
drastically for increasing $M$. In order to visualize the deviation from the
SM, it is useful to normalize the coupling $g_{h\overline{t}t}$ to its SM
expectation $\sqrt{2} \,m_t/v_u$. To this end, we define
\be
\gamma_{h\overline{t}t} ~=~ \frac{g_{h\overline{t}t} \,v_u}{\sqrt{2}\; m_t} \ .
\ee
The plot in Figure~\ref{fig:httvM} shows that the deviation of this quantity
from its SM value is of the order of a 
few percent for small values of $M$. Experimentally, the Higgs-top-top
coupling is however only poorly known with an uncertainty of about
30~percent~\cite{Giardino:2013bma,Lopez-Val:2013yba,Corbett:2015ksa}.  
The correction to $\gamma_{h\overline{t}t}$ is therefore expected to be more
important for precision observables, where the top quark and the Higgs appear
in loops so that indirect constraints apply.

Searching for correlations, we find that $\gamma_{h\overline{t}t}$ and $|V_{tb}|$
show a similar $M$-dependence for the points corresponding to the second
(orange) solution of Eq.~\eqref{eq:GRVInversion},
cf. Figure~\ref{fig:Vtbvhtt}. Such a correlation can be of phenomenological
interest as the top quark decays nearly instantaneously into a bottom quark
and a $W$ boson. For smaller values of~$M$, it thus amplifies the reduction of
associated Higgs production, $gg\rightarrow \overline{t}tH$, a process
currently searched for at the LHC \cite{Khachatryan:2014qaa, Aad:2015gra}.
 In contrast to this, the first (red) solution of
Eq.~\eqref{eq:GRVInversion} does not feature such a correlation. 

Besides the reduction of the normalized coupling $\gamma_{h\overline{q}q}$, we
also obtain flavour changing couplings of the Higgs boson which can be relevant for
flavour changing neutral currents (FCNC). The corresponding weak experimental
bounds for heavy quarks will eventually become stricter. A detailed analysis
of FCNC effects, including the $Z$ boson coupling, is therefore well motivated
but beyond the scope of this paper. Concerning the light quarks, for
which strict bounds on FCNC exist, we stress that these are naturally
suppressed in our PS flavour model.




\section{\label{sec:conclus}Conclusion}
\cleqn

Combining the idea of vertical and horizontal unification, we have
constructed a Pati-Salam symmetric model with gauged $SU(3)_I\times SU(3)_{II}$ flavour
symmetry. In the unbroken phase, the model is characterized by an explicit
left-right (i.e.\ $Z_2$) symmetry and an electroweak symmetry breaking Higgs
bi-doublet. Analogous to the construction of Grinstein, Redi and Villadoro (GRV), the
flavour structure of the SM quarks and leptons originates in 
a renormalizable Lagrangian involving new heavy fermions as well as a
set of scalar flavon fields.
Vacuum expectation values of the latter generate masses for the additional
fermions, which in turn mediate the associated flavour symmetry breaking to
the SM fermions in a seesaw-like fashion.
In contrast to GRV, our left-right symmetric PS setup requires the introduction
of two (rather than one) heavy fermionic partners for each SM fermion. As a
consequence, the cancellation of flavour gauge anomalies entails a further
extension of the fermionic particle content. This finds a natural realization
in the lepton sector where light Majorana neutrino masses can be generated by
introducing PS neutral fermions which acquire large Majorana masses from
another set of flavon fields. These flavons break the flavour symmetry
completely at a very high scale so that all flavour gauge bosons decouple from
low-energy physics.

In the quark sector, the effective SM Yukawa matrices are related to the VEVs
of the flavon fields neither linearly as in the MFV approach nor in a simple inverse
way as in the original GRV setup. The approximate relations, obtained by
integrating out the heavy fermions and given in Eq.~\eqref{eq:Yukrelations}, do
not fix the flavon VEVs uniquely. Besides the unknown mixing of the
right-chiral quarks (which is physical in our PS model), there exists a total
of eight solutions for the flavon VEVs, all giving the same approximate SM
Yukawa matrices. Inserting these vacua into the complete fermion mass
matrices, it is possible to extract the masses and the mixing angles of the
three generations of SM quarks. Comparing these with the physical values, we
find that the approximate formulas of Eq.~\eqref{eq:Yukrelations} do not
adequately describe the third generation. Such a behaviour is expected and can
be traced back to substantial mixing of the third generation of quarks with
the corresponding heavy fermionic partners. In order to correct for this
effect, we define adapted flavour parameters and take these as input for our
numerical scan. Similarly to the GRV case, the construction of the model
guarantees that non-standard flavour transitions between the first and the
second generation are highly suppressed, while new physics effects involving
the third generation of quarks are expected to be phenomenologically
relevant. Characteristic effects include the non-unitarity of the effective
CKM matrix, the coupling of the SM right-chiral quarks to the electroweak
$W^\pm$ boson and anomalous couplings of the $Z$ as well as the Higgs boson 
to two SM quarks. Quantitatively, all these effects are of order
$\epsilon_{u,d}^2$, where $\epsilon_{u,d}=\VEV{h^0_{u,d}}/M$ denotes the ratio
of the Higgs VEV over the sole explicit mass scale~$M$ of the model. By
construction, $M$ corresponds to the scale of the lightest fermionic
partners, and we choose it to be in the TeV regime. The discussed new physics
signatures of our PS GRV setup are therefore below the current
experimental sensitivity. Yet, depending on the type of solution for the
flavon VEVs, the phenomenological predictions beyond the SM might be tested in
the near future. 

In this work we have not addressed the question of the underlying dynamics which
governs the vacuum structure of the flavon fields. In our numerical scan we find that
viable flavon VEVs feature hierarchical patterns which are more or less 
comparable to those of the inverse Yukawa matrices. Ideas of generating 
hierarchical flavon VEVs from an appropriate flavour-invariant scalar
potential have already been discussed in the context of MFV, and we intend to
apply similar methods of spontaneous flavour symmetry breaking to our PS GRV setup
in a future publication.
Another subject of further investigation concerns the lepton sector. In this work, 
we have presented a neutrino extension in which the flavour structure of the neutrinos 
decouples from the hierarchical patterns of the quark sector. The construction of a fully 
realistic charged lepton sector, including a detailed discussion of lepton-flavour violating 
processes and lepton non-universality, is left for future work.




\section*{Acknowledgements}

This work is supported by the Deutsche Forschungsgemeinschaft
(DFG) within the Research Unit FOR 1873
(Quark Flavour physics and Effective field Theories).  
FH acknowledges support from the Naturwissen\-schaftlich-Technische 
Fakult\"at, University of Siegen.




\section*{Appendix}

\begin{appendix}

\section{Lepton sector}
\cleqn


\subsection{Neutrinos}
\label{app:leptons}

In order to accommodate Majorana neutrinos,
we have extended the Pati-Salam GRV model by $\mathcal L^\nu_\text{Yuk}$ of
Eq.~\eqref{eq:fermionlagrangian-nu}. Integrating out the Majorana
fermions $\overline \Theta_L$ and $\Theta_R$ generates an effective heavy
Majorana mass for $\Sigma_R^\nu$, cf. Eq.~\eqref{eq:effnuL}. This induces
a Majorana mass term for the left-chiral neutrino $\overline q_L^\nu$ via the
terms of Eq.~\eqref{eq:fermionlagrangian}. In the first part of this appendix,
we derive the structure of the resulting neutrino mass matrix. In the second
part, we discuss an alternative ansatz in which $q_R^\nu$ (rather than
$\Sigma_R^\nu$) acquires a heavy Majorana mass. This is one of several
alternative examples leading to light neutrino masss via
Eq.~\eqref{eq:fermionlagrangian}. The flavour structure of the resulting mass
matrix is, however, less attractive as will be discussed exemplarily below.


\subsubsection{Preferred neutrino extension}

Starting with the neutral part of the Lagrangian of Eq.~\eqref{eq:lagrangian},
it is straightforward to integrate out the heavy fields $\Theta_{L,R}$ and
$\Xi_{L.R}$. Inserting the flavon and Higgs VEVs we find
\begin{align}
     \mathcal{L}^{\nu}_{\text{mass}}~ = ~
     &\phantom{+}\;
     \lambda\;\overline{q}^\nu_{L}\,\mbox{$\frac{v_u}{\sqrt{2}}$}\,\Sigma^\nu_{R} 
     + M\,\overline{\Sigma}^\nu_L\,t_u\,\Sigma^\nu_R
     + M \,\overline{\Sigma}^\nu_L\, q^\nu_{R} \;+h.c. \nonumber \\
     &-\lambda\, \mbox{$\frac{v_u}{\sqrt{2}}$}\;\overline{q}^\nu_{L}\,s^{-1}\,q^\nu_R \;+h.c. \nonumber\\
     &-\,\mbox{$\frac{1}{2}$}
     ~ \Sigma^\nu_R \, M_{\Sigma_R^\nu} \, \Sigma^\nu_R \;+h.c. \, ,\label{eq:numassequation}
  \end{align}
with $M_{\Sigma_R^\nu} =
\frac{(\varphi_\alpha'   \Lambda_\varphi)^2}{\Lambda_\nu} \,s_\nu^{-1}$, 
cf. Eq.~\eqref{eq:effnuL}. To arrive at this simple intermediate result, we
have assumed $\frac{M^2}{\Lambda'_\nu} \ll \Lambda_\nu$.
Next, we integrate out $\Sigma^\nu_R$ to obtain
\begin{align}
     \mathcal{L}^{\nu}_{\text{mass}} ~= ~&
\frac{\Lambda_\nu}{2\,(\varphi'_\alpha\,\Lambda_\varphi)^2}
\left(\lambda\, \mbox{$\frac{v_u}{\sqrt{2}}$}\, \overline q_L^\nu+ M \, \overline \Sigma_L^\nu t_u  \right) 
s_\nu 
\left(\lambda\,  \mbox{$\frac{v_u}{\sqrt{2}}$}\, \overline q_L^\nu+ M\,\overline \Sigma_L^\nu t_u  \right)^T
     \,+\, h.c. 
\nonumber \\
     &+ M \,\overline{\Sigma}^\nu_L\, q^\nu_{R} - \lambda\, \mbox{$\frac{v_u}{\sqrt{2}}$}\;\overline{q}^\nu_{L}\,s^{-1}\,q^\nu_R  \;+h.c. \, ,
  \end{align}
where we have separated the Majorana mass term in the first line from the
Dirac terms in the second line. $\overline \Sigma^\nu_L$ and $q^\nu_R$ form a
Dirac pair and can be integrated out simultaneously. It is interesting to note
that already the equation of motion of $q^\nu_R$ is sufficient to deduce the
effective light neutrino mass term. Inserting $\overline{\Sigma}^\nu_L \rightarrow
\lambda\,\frac{v_u}{\sqrt{2}\,M}\,\overline{q}^\nu_{L}\,s^{-1}$ yields
\be    \label{eq:effneutrinomass-app} 
\mathcal{L}^{\nu}_{\text{mass}} ~=~
{\mbox{$\frac{1}{2}$}} \, 
\overline q_L^\nu  \,
\left[
{\mbox{$\frac{\Lambda_\nu \,v_u^2\,\lambda^2}{2\,(\varphi_\alpha' \Lambda_\varphi)^2}$}}
 \left(\id +
  s^{-1}t_u\right)\,s_\nu\,\left(\id + s^{-1}t_u\right)^T \right]
\, {\overline q_L^\nu}^T \ ,
\ee
which corresponds  to the effective mass matrix reported in
Eq.~\eqref{eq:effneutrinomass}.  Following this derivation, the occurrence of
the second term in the sum $(\mathbb 1 +s^{-1} t_u)$ is traced back to the
coupling of $\Sigma_R^\nu$ to $\overline q_L^\nu$  via $\overline
\Sigma^\nu_L$ and $q_R^\nu$, cf. Eq.~\eqref{eq:numassequation}.

It is
important to note that a high scale VEV $\VEV{S'_\nu}$ is essential. A
vanishing or too small value would imply only a small Majorana mass for
$\Theta_R$ as well as a mixing with $\Sigma^\nu_R$.  Integrating out the
heavy fields $\Sigma^\nu_R$, $\overline{\Sigma}^\nu_L$ and $q^\nu_R$ results
in a strongly hierarchical Dirac mass term for $\overline{q}^\nu_L$ and
$\Theta_R$ which is phenomenologically excluded, unless the Majorana
contribution, identical to Eq.~\eqref{eq:effneutrinomass}, to the neutrino
mass dominates.


\subsubsection{An alternative neutrino extension}

Altering the flavour quantum numbers of $\Phi$~\&~$\Phi'$ from
${\bf{(8,1)}}$~\&~${\bf{(1,8)}}$ to  
${\bf{(3,\overline 3)}}$~\&~${\bf{(3,\overline 3)}}$, entails a slightly
modified form of $\mathcal L^\nu_\text{Yuk}$ in
Eq.~\eqref{eq:fermionlagrangian-nu}: the $Z_2$ pair 
$(\overline \Xi_L,\Sigma_R)$ simply gets replaced by 
 $(\overline q_L, q_R)$. Analogous to the preferred case, one obtains a
Majorana mass term for $q^\nu_R$ after integrating out $\Theta_{L,R}$.
\begin{align}
     \mathcal{L}^{\nu}_{\text{mass}}~ = ~
     &\phantom{+}\;
     \lambda\;\overline{q}^\nu_{L}\,\mbox{$\frac{v_u}{\sqrt{2}}$}\,\Sigma^\nu_{R} 
     + M\,\overline{\Sigma}^\nu_L\,t_u\,\Sigma^\nu_R
     + M \,\overline{\Sigma}^\nu_L\, q^\nu_{R} \;+h.c. \nonumber \\
     &-\lambda\,
     \mbox{$\frac{v_u}{\sqrt{2}}$}\;\overline{q}^\nu_{L}\,s^{-1}\,q^\nu_R
     \;+h.c. \nonumber \\
     &-\,\mbox{$\frac{1}{2}$}
     ~ q^\nu_R \, M_{q_R^\nu} \, q^\nu_R \;+h.c. \, ,\label{eq:numassequation-alt}
  \end{align}
where $M_{q_R^\nu} =\frac{\Lambda_\varphi^2}{\Lambda_\nu}  \,
 {\varphi'}^T\, s_\nu^{-1}\,{\varphi'}$. The heavy right-chiral neutrino
 $q_R^\nu$ couples to $\overline q_L^\nu$ both directly and indirectly via
 $\overline \Sigma_L^\nu$ and $\Sigma_R^\nu$, thereby generating an effective
 Majorana mass for the left-chiral neutrino $\overline q_L^\nu$. Integrating
 out $q_R^\nu$ and $\Sigma^\nu_{L,R}$ in Eq.~\eqref{eq:numassequation-alt} results
 in 
\be    \label{eq:badnumass}
\mathcal{L}^{\nu}_{\text{mass}} ~=~
{\mbox{$\frac{1}{2}$}} \, \overline q_L^\nu  \, \left[
{\mbox{$\frac{\Lambda_\nu \,v_u^2\,\lambda^2}{2\, \Lambda_\varphi^2}$}}
 (t_u^{-1} + s_{}^{-1})\,{\varphi'}^{-1}\, s_\nu\,{{\varphi'}^{-1}}^{T} \,(t_u^{-1} + s_{}^{-1})^T
\right] \,
{\overline q_L^\nu}^T \ .
\ee
Comparing Eq.~\eqref{eq:badnumass} and Eq.~\eqref{eq:effneutrinomass-app} we
notice two important differences. First, due to its transformation as a
${\bf{(3,\overline 3)}}$ in flavour space, the matrix $\varphi'$ contributes
non-trivially to the flavour structure of the light neutrinos. Second, the
factor $(t_u^{-1} + s_{}^{-1})\sim Y_u$ features strong hierarchies in contrast
to  $(\id + s^{-1}\,t_u)$. These hierarchies must be approximately
compensated by an even stronger hierarchy in $M_{q_R^\nu}^{-1}$ in order to
generate a phenomenologically viable neutrino mass spectrum. As such a
cancellation of hierarchies appears rather ad~hoc, we abandon this
alternative neutrino extension.


\subsection{Charged leptons}
\label{app:alt_chargedleptons}

As already mentioned in Section~\ref{sec:approx}, the structure of the charged
lepton Yukawa matrix is identical to the one of the down-type quarks. At the
GUT scale we have $Y_e = Y_d$ which is a reasonable approximation. Deviations
from this equality can be induced by allowing the flavon fields $S$ or $T^{(\prime)}$
to transform non-trivially under $SU(4)$. Due to their appearance in the
Lagrangian of Eq.~\eqref{eq:fermionlagrangian}, the only alternative
representation is the adjoint ${\bf{15}}$ of $SU(4)$. Such a choice does not
alter the quark sector, but it introduces a Georgi-Jarlskog factor of $-3$ for
the leptons~\cite{Georgi:1979df}. For instance, if $S$ is kept in the singlet
representation of $SU(4)$ while $T^{(\prime)}$ furnishes the adjoint
representation, we obtain a difference between the Yukawa matrices of the
charged leptons and down-type quarks. Explicitly, and adopting the
approximation of Section~\ref{sec:approx}, we obtain
\be
 Y_d ~\sim ~ \bigl(s - t'\bigr)^{-1} + s^{-1} \ ,\qquad
  Y_e ~\sim ~ \bigl(s +3\,t'\bigr)^{-1} + s^{-1} \  .
\ee
More complicated relations are possible if we allow the flavons $S$ and/or
$T^{(\prime)}$ to appear in both $SU(4)$ representations. We conclude 
that the model can therefore accommodate a viable charged lepton sector as
well.




\section[Determining~s~and~t'~numerically]{\label{sec:seesaw}Determining $\boldsymbol{s}$  and $\boldsymbol{t'}$
numerically}
\cleqn

As already discussed in Section~\ref{sec:approx}, it is a not-trivial task to
invert the relations between the
flavour breaking VEVs and the Yukawa matrices. In this Appendix, we present a
procedure for calculating $s$ and $t'$ numerically for a given pair of $Y_u$
and $Y_d$ assuming Eq.~\eqref{eq:Yukrelations} to hold exactly. We therefore write
\begin{align}
 -\tfrac{1}{\lambda}\,Y_u = (s+t')^{-1} + s^{-1} \ , \qquad
 -\tfrac{1}{\lambda}\,Y_d = (s-t')^{-1} + s^{-1} \,.
\end{align}
Defining $  H \equiv s^{-1}\,t'$ we obtain the expressions
\be
  -\tfrac{1}{\lambda}\,Y_u\,s = (\id+H)^{-1} + \id \ , \qquad
\label{eq:YuSformH}
 -\tfrac{1}{\lambda}\, Y_d\,s = (\id-H)^{-1} + \id \ ,
\ee
which can be combined to remove the $s$-dependence
\begin{align}
  \label{eq:triangularrelation} 
 G\equiv Y^{}_d \, Y_u^{-1}~=~\left[ (\id-H)^{-1} + \id \right] \left[ (\id+H)^{-1} + \id
  \right]^{-1} \ .
\end{align}
Let us assume for the moment that $G$ is a triangular matrix, which can always
be achieved by a suitable similarity transformation. 
In that case, also $H$ has a triangular form, so that
Eq.~\eqref{eq:triangularrelation} can be solved explicitly for the six
elements of $H$. We thus obtain an explicit solution for $H(G)$ in the special
case of a triangular matrix $G$. 

As $Y_d\,Y_u^{-1}$ is generally not given in a triangular form, we need to
generalize the solution for $H(G)$ to generic forms of $G$.
For this purpose we expand the function $H(G)$ as a series in the matrix $G$.
Due to the Cayley-Hamilton theorem this series stops after the quadratic 
term.\footnote{In general the theorem states that for each $n\times n$ matrix
  $M$, $M^n$ can be expressed as a polynomial $\sum_{i<n} x_i\,M^i$, where the
  factors $x_i$ depend solely on traces of powers of $M$ up to order
  $M^n$. For a detailed discussion see
  e.g.~\cite{ZhangBook,PetersenBook}.}
Hence, we can express $H(G)$ in a basis independent way by
\begin{align}\label{eq:HofGtraces}
 H(G) = a\,\id + b\,G + c\,G^2\ ,
\end{align}
where the coefficients $a$, $b$, and $c$ depend only on the traces of $G$,
$G^2$ and $G^3$. In the special case of a triangular matrix $G$, it is
possible to solve Eq.~\eqref{eq:HofGtraces} analytically for $a$, $b$ and $c$.
As these coefficients are basis independent, 
Eq.~\eqref{eq:HofGtraces} holds true in any basis; in particular
$G=Y_d\,Y_u^{-1}$ need not be of triangular form.

In principle, we can determine $H$ for a given pair of Yukawa matrices
$Y_{u,d}$ analytically. Due to their enormous length, the resulting
expressions are however not particularly instructive which is why we content ourselves
with a numerical evaluation. Having determined the matrix $H$, we can use
Eq.~\eqref{eq:YuSformH} to calculate $s$
\begin{align}
s = -\,\lambda\,Y_u^{-1} \left(\left(\id + H\right)^{-1} + \id \right)\ ,
\end{align}
and the definition of $H$ to determine $t'=s\,H$.
Thus, we have obtained numerical matrices $s$ and $t'$ for a given pair of
Yukawa matrices $Y_{u,d}$.




\section{\label{app:sequence}Diagonalizing the quark mass matrices}
\cleqn

In this appendix, we present the technical details related to the
diagonalization of the quark mass matrices, discussed in
Section~\ref{subsec:sequence}. Starting from the original basis given in
Eqs.~(\ref{eq:defPSI},\ref{eq:Mu}), we perform a sequence of basis
transformations to diagonalize $\mathcal M^u$ to second order in $\epsilon_u$.
The individual intermediate bases are labelled by a subscript ${(i)}$.

\begin{enumerate}

\item {Basis with diagonal ${s=\hat s}$:}

Thanks to the flavour symmetry $SU(3)_I\times SU(3)_{II}$ 
we can always choose a basis in which the matrix $s$ is diagonal. $\mathcal
M^u_{(1)}$  is therefore identical to $\mathcal M^u$ of Eq.~\eqref{eq:Mu} with
$s\rightarrow \hat s$. Note that this choice of basis is valid for both the
up and the down sector.


\item {Diagonalizing ${t_{u}}$:} 

In the second step, we apply the basis transformation
\be
{\overline \Psi^u_L}_{(2)} \equiv {\overline \Psi^u_L}_{(1)}
\, \text{diag}(\id,V_u^\dagger,\id) \ , \qquad
{\Psi^u_R}_{(2)} \equiv \text{diag}(V_u,U_u,\id) \, {\Psi^u_R}_{(1)} \ ,
\label{eq:psi2}
\ee
to render $t_u$ diagonal, see Eq.~\eqref{eq:diag-tu}. The resulting mass
matrix takes the form 
\be
\mathcal M^u_{(2)} ~=~ 
\begin{pmatrix}
0 &U_u^\dagger \lambda \epsilon_u  & \id \\
\id  &  \hat t_u  & 0  \\
 V_u^\dagger \lambda \epsilon_u  & 0 & \hat s
\end{pmatrix} M \ .
\ee
Notice that the unitary rotations $V_u$ and $U_u$ reappear at order
$\epsilon_u$ only. It is furthermore important to realize that the 
corresponding matrices $V_d$ and $U_d$ for the down sector differ from those
of the up sector.


\item {Diagonalizing $\mathcal M^u_{(2)}$ for $\epsilon_u=0$:}

Setting $\epsilon_u$ to zero, the mixing of the three generations disappears
completely. Introducing the following cosines and sines,
\be
   c^i_s \!= \frac{\hat{s}_i}{\sqrt{1+{\hat s_i}^2}}  \ ,  \quad
   s^i_s \!= \frac{1}{\sqrt{1+{\hat s_i}^2}} \ , 
   \quad  \qquad
   c^i_{t_u} \!= \frac{\hat{t}^i_u}{\sqrt{1+\!\!\phantom({\hat{t}}^i_u\phantom)\!\!^2}}
    \ , \quad
   s^i_{t_u} \!= \frac{1}{\sqrt{1+\!\!\phantom({\hat{t}}^i_u\phantom)\!\!^2}}
      \ ,
\ee
we can define the diagonal $3\times 3$ matrices
\be
\hat c_x = \text{diag} (c_x^1,c_x^2,c_x^3) \ , \qquad
\hat s_x = \text{diag} (s_x^1,s_x^2,s_x^3) \ ,
\ee
where the index $x$ stands for $s$ or $t_u$. With this notation at hand, 
it is straightforward to diagonalize $\mathcal M^u_{(2)}$ for
$\epsilon_u=0$. In the basis of 
\be
{\overline \Psi^u_L}_{(3)} \equiv {\overline \Psi^u_L}_{(2)}
\begin{pmatrix} 
\hat c_s & 0 &  \hat s_s \\
0&\id&0 \\
-\hat s_s & 0 & \hat c_s 
\end{pmatrix}
\ , \qquad
{\Psi^u_R}_{(3)} \equiv 
\begin{pmatrix}
\hat c_{t_u} & -\hat s_{t_u} & 0 \\
\hat s_{t_u} &\hat c_{t_u} & 0\\
0&0&\id 
\end{pmatrix} 
{\Psi^u_R}_{(2)} \ ,
\label{eq:psi3}
\ee
the full mass matrix, including the terms proportional to $\epsilon_u$, is
given by 
\be
\mathcal M^u_{(3)} ~=~ 
\begin{pmatrix} 
 (-\hat  s_s V^\dagger_{u} \hat c_{t_u}-\hat c_s U_{u}^\dagger \hat s_{t_u}) \lambda\,\epsilon_u &
    (-\hat s_s V_{u}^\dagger \hat s_{t_u} + \hat c_s U^\dagger_{u} \hat c_{t_u}) \lambda\,\epsilon_u &
    0 \\
    0 & \hat s^{-1}_{t_u} & 0 \\
    ( \hat c_s V^\dagger_{u} \hat c_{t_u}-\hat s_s U_{u}^\dagger \hat s_{t_u} ) \lambda\,\epsilon_u &
    (\hat c_s V_{u}^\dagger \hat s_{t_u} + \hat s_s U^\dagger_{u} \hat c_{t_u}) \lambda\,\epsilon_u &
    \hat s^{-1}_{s} 
\end{pmatrix} M \ .
\label{eq:Mv3}
\ee
Identifying this with the matrix of Eq.~\eqref{eq:Mu3}, defines the $3\times
3$ matrices $a_u$, $b_u$, $c_u$, $d_u$, $\hat e_u$, and $\hat f$. The latter
two are diagonal and positive definite.


\item {Block-diagonalizing $\mathcal M^u_{(3)}$ up to order $\epsilon_u^2$:}

Having parameterized $\mathcal M^u_{(3)}$ as in Eq.~\eqref{eq:Mu3}, 
we proceed to block-diagonalize this matrix up to order $\epsilon_u^2$. To
this end, we adopt the following unitary transformations,
\begin{subequations}
\bea
{\overline \Psi^u_L}_{(4)} &\equiv & {\overline \Psi^u_L}_{(3)}
 \left[\mathcal R_{12} (\xi^{u}_{12})\right]^\dagger
\left[\mathcal R_{23} (\xi^{u}_{23})\right]^\dagger
\left[\mathcal R_{13} (\xi^{u}_{13})\right]^\dagger
 \ , \\[1mm]
{\Psi^u_R}_{(4)} &\equiv& 
\left[ \mathcal R_{12} (\zeta^{u}_{12})\right]\,
\left[ \mathcal R_{23} (\zeta^{u}_{23})\right]\,
\left[ \mathcal R_{13} (\zeta^{u}_{13})\right]
 \, {\Psi^u_R}_{(3)} \ ,
\eea
\end{subequations}
where $ \mathcal{R}_{\alpha\beta}(\xi)$ denotes a ``rotation in the
$\alpha$-$\beta$ plane'', expanded to second order in~$\xi$. 
For $(\alpha,\beta)=(1,2)$, one has
\begin{align}
  \mathcal{R}_{12}(\xi) = \left(\begin{array}{ccc} \id - \tfrac12\,
\xi\,\xi^\dagger & -\xi & 0\\ \xi^\dagger & \id - \tfrac12\,\xi^\dagger\,\xi & 0 \\
0&0&\id \end{array} 
\right) \, ,
\end{align}
and the expressions for the other two pairs, $(2,3)$ and $(1,3)$, are
identical up to obvious permutations of rows and columns.
The matrices $\xi$ and $\zeta$ are given in terms of the parameters of
Eq.~\eqref{eq:Mu3} by 
\begin{subequations}
\begin{align}
{\xi}^{u}_{12} &= b_u \hat e_u^{-1} \, \epsilon_u \ ,
&\left[{\xi}^{u}_{23}\right]_{ij} &= 
\frac{-\hat e_u^i  {d^\dagger_u}^{ij}}
{\phantom(\!\!\hat e_u^i\phantom)\!\!^2 -\phantom(\!\!\hat f_j\phantom)\!\!^2} 
\,\epsilon_u \ ,
&{\xi}^{u}_{13} &= a_u c_u^\dagger  \hat f^{-2}\,\epsilon_u^2 \ ,
\\
{\zeta}^{u}_{12} &= a_u^\dagger b_u  \hat e_u^{-2} \, \epsilon_u^2 \ ,
&\left[{\zeta}^{u}_{23}\right]_{ij} &= 
\frac{-  {d^\dagger_u}^{ij}\hat f_j}
{\phantom(\!\!\hat e_u^i\phantom)\!\!^2 -\phantom(\!\!\hat f_j\phantom)\!\!^2} 
\,\epsilon_u \ ,
&{\zeta}^{u}_{13} &= c_u^\dagger  \hat f^{-1}\,\epsilon_u \ .
\end{align}
\end{subequations}
We note that the matrices ${\xi}^{u}_{23}$ and ${\zeta}^{u}_{23}$ cannot be
written as a simple product of matrices. With a little bit of algebra, it is
possible to show that the resulting mass matrix $\mathcal M^u_{(4)} $ is given
by the matrix in Eq.~\eqref{eq:Mu4}.


\item {The approximate mass basis:}

The final step of the sequence of basis transformations consists in the
diagonalization of the upper left $3\times 3$ block of $\mathcal M^u_{(4)}$,
see Eq.~\eqref{eq:Yhat}. The corresponding transformation reads
\be
{\overline \Psi^u_L}_{(5)} \equiv {\overline \Psi^u_L}_{(4)}
\, \text{diag}(\mathcal V_u^\dagger,\id,\id) \ , \qquad
{\Psi^u_R}_{(5)} \equiv \text{diag}(\mathcal U_u,\id,\id) \, {\Psi^u_R}_{(4)}
\ .
\label{eq:psi5}
\ee
The basis ${\Psi^u_{L,R}}_{(5)}$ corresponds to the approximate mass basis
denoted by ${\Psi'^u_{L,R}}$ in Section~\ref{sec:GRVQuarkFlavour}.

\end{enumerate}

\noindent Applying this sequence of basis transformations to the gauge-kinetic terms
changes the $\mathcal K$ matrices of Eq.~\eqref{eq:GRVbrokenkinlag} to
\begin{subequations}\label{eq:gaugekinflavour}
\footnotesize
\begin{align}
{\mathcal K^+_{L}}_{(5)} &=
\begin{pmatrix}
\mathcal{V}_u\mathcal{V}^\dagger_d
-\mathcal{V}_u\frac{1}{2} \Bigr(\xi^u_{12}{\xi^u_{12}}^\dagger +\xi^d_{12}{\xi^d_{12}}^\dagger\Bigr)\mathcal{V}^\dagger_d & \mathcal{V}_u\,\xi^d_{12} & -\mathcal{V}_u\Bigl(\xi^u_{13}-\xi^d_{13}-\xi^d_{12} \xi^d_{23} \Bigr) 
\\
{\xi^u_{12}}^\dagger \mathcal{V}^\dagger_d & \!\!\!{\xi^u_{12}}^\dagger \xi^d_{12}+ \xi^u_{23} {\xi^d_{23}}^\dagger \!\!\! & -\xi^u_{23} 
\\
\Bigl({\xi^u_{13}}^\dagger - {\xi^d_{13}}^\dagger + {\xi^u_{23}}^\dagger
     {\xi^u_{12}}^\dagger \Bigr)\mathcal{V}^\dagger_d & -{\xi^d_{23}}^\dagger
     & \id - \frac{1}{2} \Bigr({\xi^u_{23}}^\dagger\xi^u_{23}+{\xi^d_{23}}^\dagger\xi^d_{23}\Bigr) 
\end{pmatrix}\!\!+\:\!\!\mathcal O(\epsilon^3_{u,d})\ , \\[1ex]
{\mathcal K^+_{R}}_{(5)} &=
\begin{pmatrix}
\mathcal{U}_u\,\zeta^u_{13}{\zeta^d_{13}}^\dagger\,\mathcal{U}_d^\dagger & \mathcal{U}_u\,\zeta^u_{13}{\zeta^d_{23}}^\dagger & -\mathcal{U}_u\,\zeta^u_{13} 
\\
\zeta^u_{23}{\zeta^d_{13}}^\dagger\,\mathcal{U}_d^\dagger & \zeta^u_{23}{\zeta^d_{23}}^\dagger & -\zeta^u_{23} 
\\
-{\zeta^d_{13}}^\dagger\,\mathcal{U}_d^\dagger & -{\zeta^d_{23}}^\dagger & \id -\frac{1}{2} ({\zeta^u_{13}}^\dagger\zeta^u_{13}+{\zeta^u_{23}}^\dagger\zeta^u_{23} +{\zeta^d_{13}}^\dagger\zeta^d_{13}+{\zeta^d_{23}}^\dagger\zeta^d_{23}) 
\end{pmatrix} + \,\mathcal O(\epsilon^3_{u,d})\ , 
\displaybreak[0]\\[1ex]
{\mathcal K'^0_{L}}_{(5)} &=
\begin{pmatrix}
\mathcal{V}\,\xi_{12}{\xi_{12}}^\dagger\,\mathcal{V}^\dagger & -\mathcal{V}\,\xi_{12} & -\mathcal{V}\,\xi_{12} \xi_{23} \\
-{\xi_{12}}^\dagger\,\mathcal{V}^\dagger & \id-{\xi_{12}}^\dagger \xi_{12}- \xi_{23} {\xi_{23}}^\dagger \!\! & \xi_{23}  \\
-{\xi_{23}}^\dagger {\xi_{12}}^\dagger\,\mathcal{V}^\dagger & {\xi_{23}}^\dagger &  {\xi_{23}}^\dagger\xi_{23}
\end{pmatrix}\!\!+\:\!\!\mathcal O(\epsilon^3)\ , \\[3ex]
{\mathcal K^0_{R}}_{(5)} &=
\begin{pmatrix}
\mathcal{U}\,\zeta_{13}{\zeta_{13}}^\dagger\,\mathcal{U}^\dagger & \mathcal{U}\,\zeta_{13}{\zeta_{23}}^\dagger & -\mathcal{U}\,\zeta_{13}  \\
\zeta_{23}{\zeta_{13}}^\dagger\,\mathcal{U}^\dagger & \zeta_{23}{\zeta_{23}}^\dagger & -\zeta_{23}  \\
-{\zeta_{13}}^\dagger\,\mathcal{U}^\dagger & -{\zeta_{23}}^\dagger & \id - {\zeta_{13}}^\dagger\zeta_{13}+{\zeta_{23}}^\dagger\zeta_{23} 
\end{pmatrix} + \,\mathcal O(\epsilon^3)\ .
\end{align}
\end{subequations}
As the basis transformations depend on the isospin components, we have
labelled the resulting flavour structures by superscripts which indicate
whether they relate to  charged or neutral currents. In the latter case, we
have suppressed the isospin indices for convenience. Note that ${\mathcal
  K^0_{R}}_{(5)}$ can be derived from ${\mathcal   K^+_{R}}_{(5)}$ by dropping
the indices $u$~and~$d$.

\end{appendix}




\end{document}